\documentclass[useAMS, usenatbib]{mn2e} 
\usepackage{aas_macros}
\usepackage{graphicx}
\usepackage{subfig}
\pdfoutput = 1
\usepackage{amsmath}
\usepackage{mathtools}
\usepackage{pdflscape}\usepackage{longtable}
\usepackage{breqn}
\title{Long-Baseline Interferometric Multiplicity Survey of the Sco-Cen OB Association}
\author[A.C. Rizzuto et al.]{\parbox{\textwidth}{A.C. Rizzuto$^{1}$, M.J. Ireland$^{1,3}$, J.G. Robertson$^{2}$,  Y. Kok$^2$,  P.G. Tuthill$^2$,  B.A. Warrington$^{1}$,  X. Haubois$^{2}$, W.J. Tango$^2$, B. Norris$^2$,  T. ten Brummelaar$^5$, A.L. Kraus$^4$, A. Jacob$^2$, C. Laliberte-Houdeville$^2$}
\\
\\
\vspace{0.1cm}
\parbox{\textwidth}{$^{1}$Department of Physics and Astronomy, Macquarie University, Sydney NSW, 2109, Australia\\
$^{2}$Sydney Institute for Astronomy (SIfA), School of Physics, University of Sydney NSW, 2006, Australia\\
$^{3}$Australian Astronomical Observatory, Epping NSW 2121, Australia\\
$^{4}$The University of Texas, Department of Astronomy, Austin, Texas 78712, USA\\
$^{5}$Center for High Angular Resolution Astronomy, Georgia State University, P.O. Box 3969, Atlanta, GA 30302, USA}}

\newcommand{\mdot}{M$_{\odot}$~}

\begin{document}

\pagerange{\pageref{firstpage}--\pageref{lastpage}} \pubyear{2013}

\maketitle

\begin{abstract}
We present the first multiplicity-dedicated long baseline optical interferometric survey of the Scorpius-Centaurus-Lupus-Crux association. We used the Sydney University Stellar Interferometer to undertake a survey for new companions to 58 Sco-Cen B-type stars and have detected 24 companions at separations ranging from 7-130\,mas, 14 of which are new detections. Furthermore, we use a Bayesian analysis and all available information in the literature to determine the multiplicity distribution of the 58 stars in our sample, showing that the companion frequency is $F=1.35\pm0.25$ and the mass ratio distribution is best described by $q^\gamma$ with $\gamma=-0.46$, agreeing with previous Sco-Cen high mass work and differing significantly from lower-mass stars in Tau-Aur. Based on our analysis, we estimate that among young B-type stars in moving groups, up to 23\% are apparently single stars. This has strong implications for the understanding of high-mass star formation, which requires angular momentum dispersal through some mechanism such as formation of multiple systems.
\end{abstract}


\section{Introduction}
\label{intro}
Multiplicity properties of recently formed stars can provide valuable insight into the the understanding of star formation mechanisms \citep{blaauw91}. For more than a decade it has been widely accepted that at least half of all solar-type stars form in pairs \citep{mathieu94}, though multiple systems are still a relatively poorly understood part of star formation. One particular unknown aspect is the role of multiplicity in the redistribution of angular momentum during star formation \citep{larson10}. Observations have also revealed that $70-90$\% of stars form in clusters \citep{lada03}.

Detailed knowledge of the multiplicity of a primordial stellar population would be the ideal. This would be a population of stars whose formation precesses have finished and which have stopped accreting gas from their surroundings, but before dynamical interactions and stellar evolution have altered the multiplicity distribution. Stellar OB associations are the closest match to these conditions, by virtue of their low density and youth, and provide a large sample of young, newly formed stars for multiplicity study.

\begin{figure}
\includegraphics[width=0.5\textwidth]{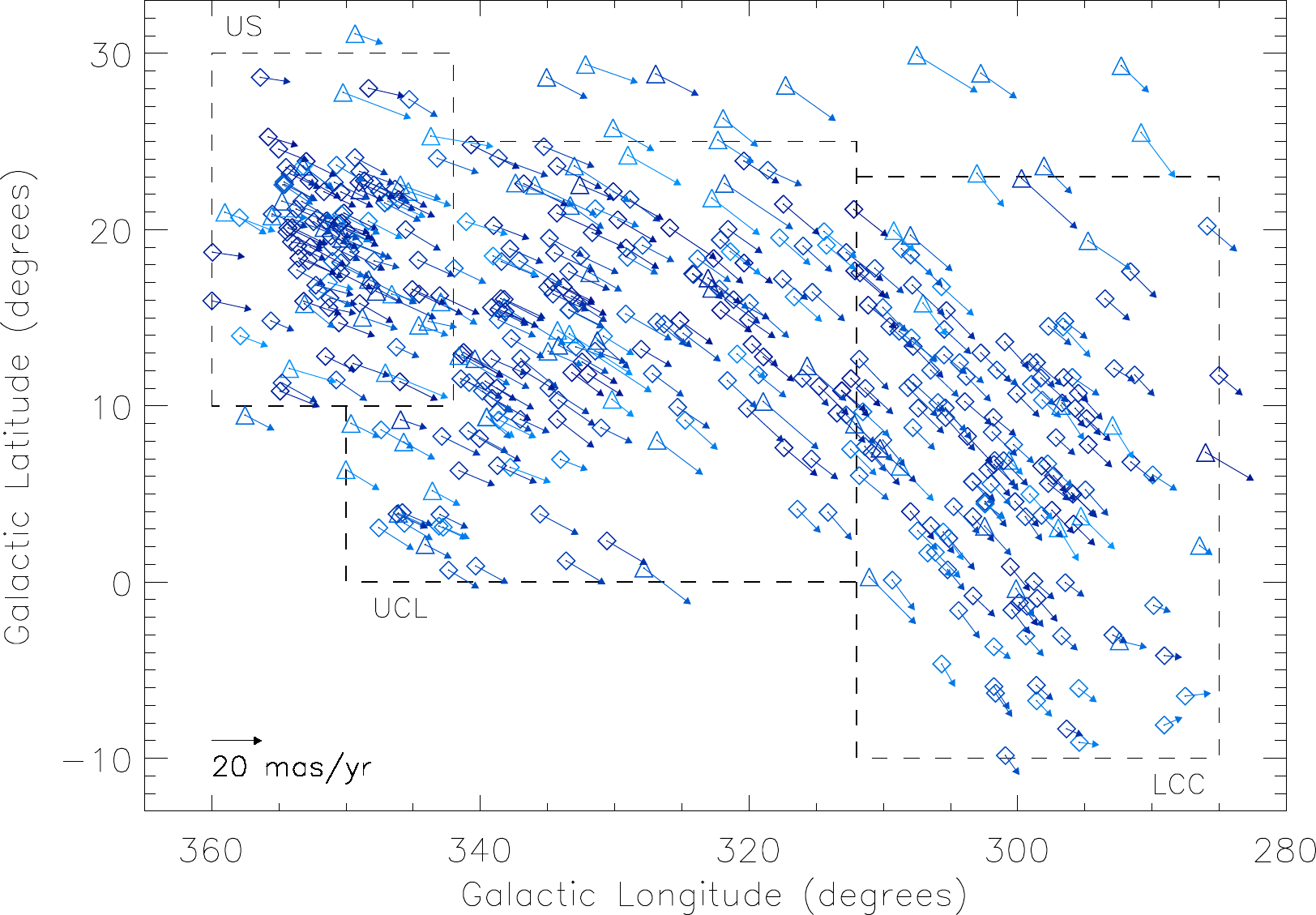}
\caption{The Sco OB2 members identified by \citet{myfirstpaper} with \emph{Hipparcos} proper motion vectors. The classical subgroups are labelled and bounded by dotted lines. Darker blue indicates higher membership probability, with a minimum of 50\% being the selection criterion for this figure.}
\label{mysco}
\end{figure}

The Scorpius-Centaurus-Lupus-Crux OB Association (Sco OB2, Sco-Cen) is the nearest region to the sun with recent massive star formation \citep{preibisch08}. The association has been classically divided into three  distinct sub-groups (see  Figure \ref{mysco}), Upper-Scorpius (US), Upper-Centaurus-Lupus (UCL), and Lower-Centaurus-Crux (LCC) \citep{blaauw46} , with mean parallaxes of 6.9, 7.1 and 8.5 milli-arcseconds respectively, or distances of 145, 143 and 118\,pc \citep{zeeuw99}. UCL and LCC have little interstellar material associated with them, whereas filamentary material can be observed towards US which is connected to the Ophiuchus cloud complex, a region of ongoing star formation \citep{geus92}. Photometry has demonstrated that the Ophiuchus cloud complex is on the near side of US at approximately 130\,pc \citep{mamajek08b}, and isochrone fitting gives mean ages for the sub-groups as ~5\,Myr for US, ~16\,Myr for UCL and ~16\,Myr for LCC \citep{geus89}. More receently, the US subgroup has been show to potentially be as old as $\sim$10\,Myr \citep{pecaut12}

In our first paper concerning the Sco-Cen OB association (Rizzuto et al. 2011), we produced an improved high-mass membership determination which  included 436 stars bluer than $B-V=0.6$. This is a large sample of stars which are as young as 5\,Myrs to survey for multiplicity information. 

Over the past decade significant progress has been made in characterising the binary population of Sco OB2. A survey of 199 A and late B-type stars in Sco OB2 was done by \citet{kouwenhoven05} using the ESO 3.6\,m telescope at La Silla, Chile with the ADONIS/SHARPII+ imaging instrument with the aim of detecting new visual binaries. They detected 74 candidate physical companions around primaries fainter than $V\sim6$ and with angular separations of 0.22" to 12.4". Of these, 41 were previously unseen. Another study by \citet{shatsky02} examined 115 B-type stars in the Sco OB2 association for visual companions using the ADONIS near-infrared coronograph on the ESO 3.6\,m telescope. This study detected 37 physical companions to Sco-Cen stars, 10 of which were new detections. 

There is also a substantial body of work concerning spectroscopic companions to high-mass stars. A large compilation of orbital parameters for single- (SB1) and double-lined (SB2) spectroscopic binaries can be  found in \citet{levato87}. More recently, there are a number of rotation and radial velocity  studies of early type stars which identify spectroscopic binaries have been published \citep{brown97,jilinski06}. 

In between the small separations of the spectroscopic binaries and the wider separations (28-1610\,AU) of the visual binaries observed by \citet{shatsky02} and \citet{kouwenhoven07} there is a range of separations of approximately 1-10\,AU which remain relatively unstudied. Binary systems with these separations include rapidly rotating and possibly pulsating B-type stars, and cover a regime in which radial velocity measurements are not possible with current instrumentation. The purpose of our study is to present a survey of the Sco OB2 association for binary separations within this niche using the Sydney University Stellar Interferometer, and to use our new observations, in conjunction with the knowledge in the literature, to determine the multiplicity properties of the young B-type stars in Sco-Cen. This will address the question of whether B-type stars form alone or as part of a double or multiple system.



\section{Observations and Data Reduction}
\subsection{Target Sample}
\begin{figure}
\subfloat[Sample Sky Positions \label{susiobs_context}]{\includegraphics[width=0.5\textwidth]{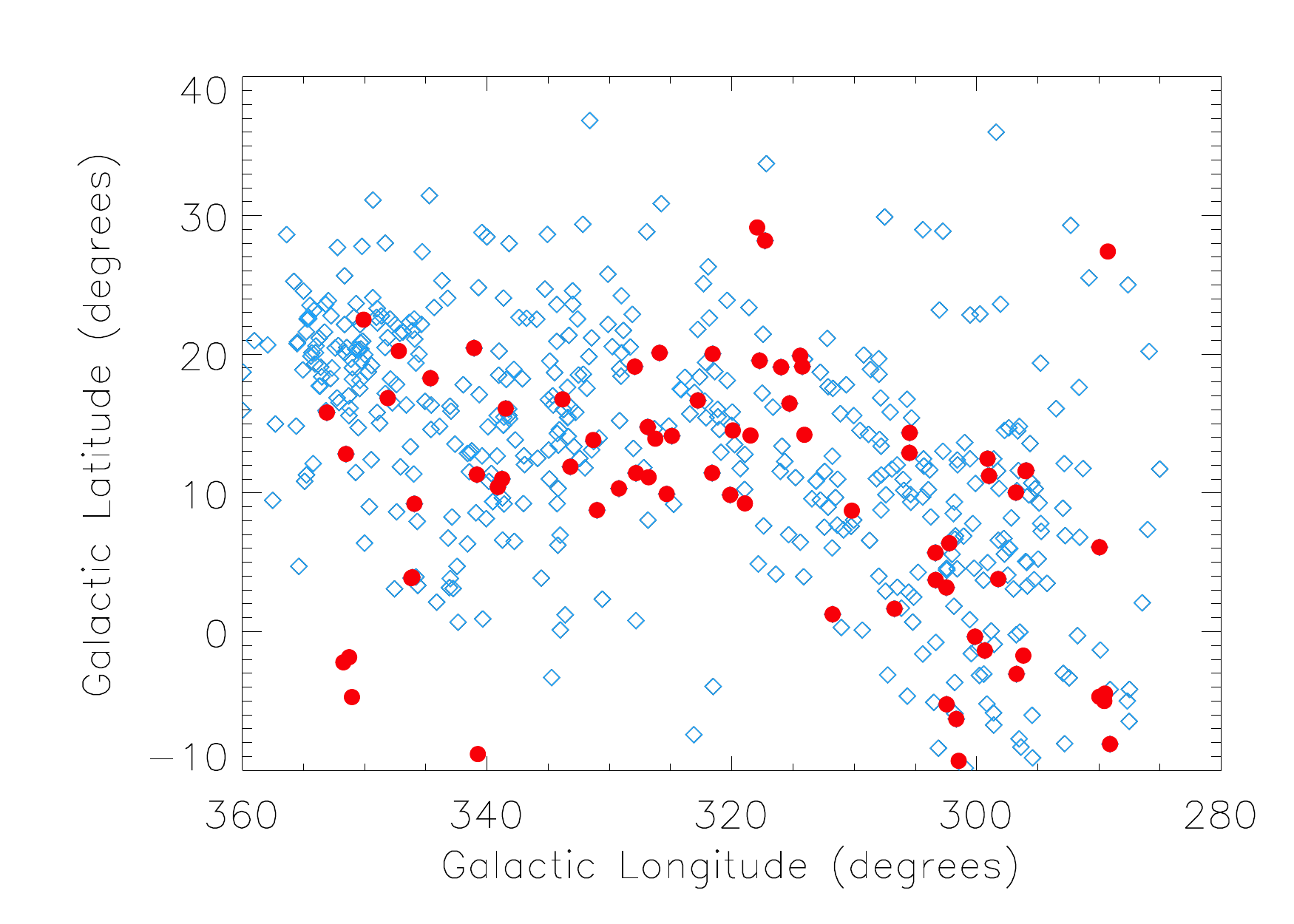}}\\
\subfloat[Sample Proper Motions \label{binary_prop}]{\includegraphics[width=0.5\textwidth]{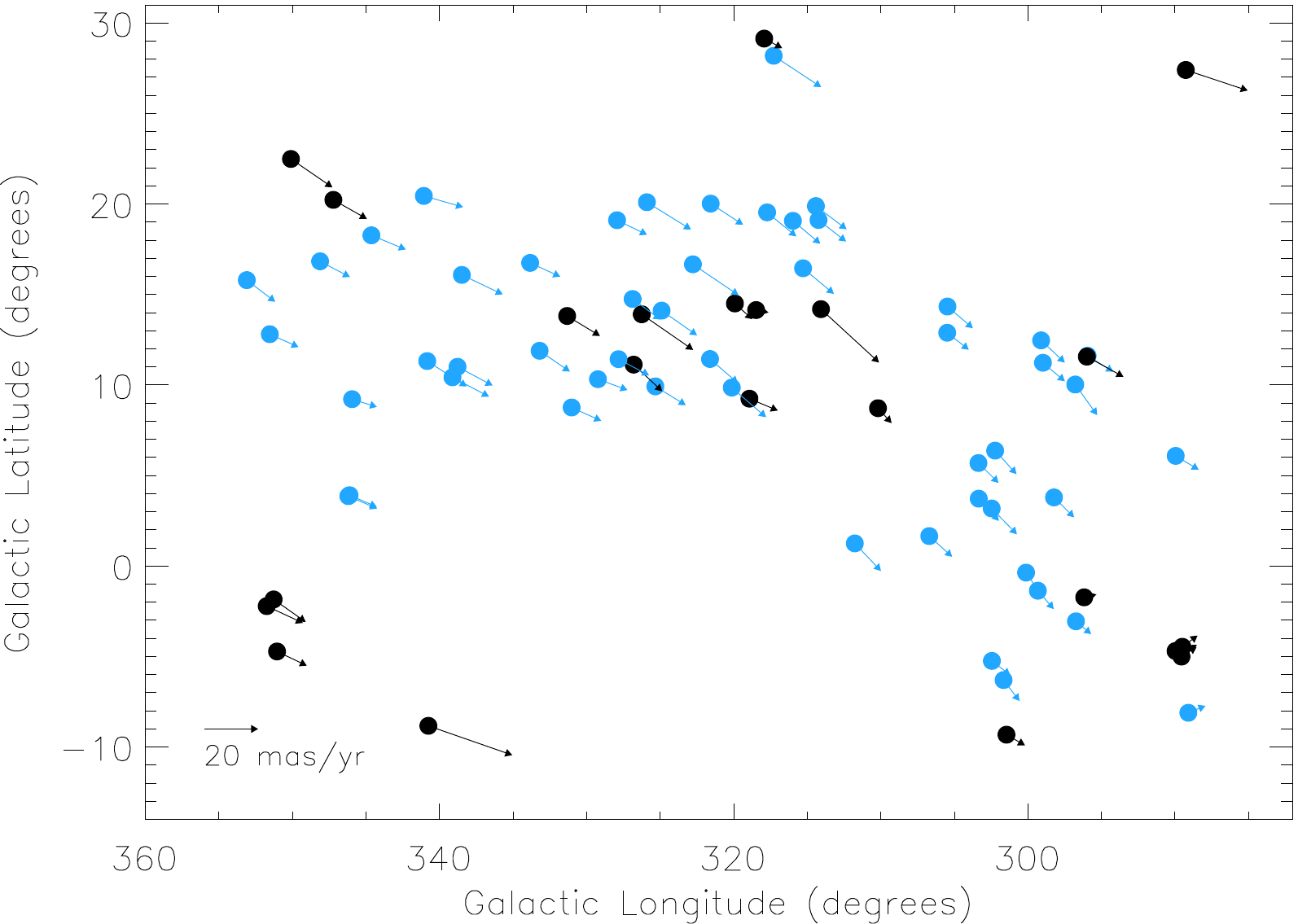}}
\caption{The on-sky locations of the Sco-Cen region high-mass stars observed in our survey. Blue squares indicate the \citet{myfirstpaper} members and red circles indicate the stars in our sample. Note the lack of high-mass stars (blue circles) outside of the Sco-Cen regions. The second figure illustrates the proper motion vectors of the stars in our sample. Blue objects once again represent members in the \citet{myfirstpaper} selection with greater than 50\% membership probability. The lack of  highly deviant proper motions highlights the possibility that multiplicity induced proper motion offsets might explain the exclusions.}
\label{samp_diag}
\end{figure}

Our aim was to observe all stars within the area of sky occupied by Sco-Cen which were brighter than $5^{th}$ apparent visual magnitude and bluer than $B-V = -0.1$ magnitudes. There are 75 stars which fit this criterion, and of these we observed 58. The spatial distribution and proper motion of our observed targets, in relation to the \citet{myfirstpaper} membership, is shown in Figure \ref{samp_diag}. The decision to observe all stars within the given colour and magnitude range, rather than just the 52 members in the \citet{myfirstpaper} selection, was motivated in two ways. Firstly, the presence of undetected binarity can affect the \emph{Hipparcos} proper motions upon which the membership determination was based \citep{zeeuw99}. \emph{Hipparcos} measurements were carried out over a period of 3.3 years. Hence, unresolved binary systems, especially those  with periods greater than 3.3 years, can affect the observed centre of motion. The typical magnitude of this error has been shown to be $\sim2$\,mas $yr^{-1}$ \citep{weilen97}, which is larger than the average \emph{Hipparcos} proper motion uncertainty. Secondly, bright, blue high-mass stars in the region of space which is considered to be Sco-Cen are young and almost certainly formed as part of the association and have since undergone dynamical changes which affect a kinematic based membership. Indeed, applying our above magnitude and colour filter to the \emph{Hipparcos} catalog bounded by $(285^\circ<l<360^\circ)$ and $(-10^\circ<b<40^\circ)$ clearly depicts a concentration of the bluest stars in the Sco-Cen subgroups and a paucity outside of these regions (Fig \ref{susiobs_context}). Similarly, Figure \ref{binary_prop} demonstrates that none of the high-mas stars in the Sco-Cen region of sky have large offsets from the expected member proper motions. Table \ref{obsed_stars} lists all stars observed with SUSI and the corresponding detection limits.

\subsection{Observations}
The observations were performed with the Sydney University Stellar Interferometer (SUSI) on a 15\,m baseline, using the PAVO beam combiner \citep{pavo08}. SUSI operates in 25 optical wavelength channels between 550 and 800\,nm, and on the 15\,m baseline has an angular resolution of 7\,mas. The coherence length of each spectral channel of the PAVO beam combiner is 30\,$\mu$m, which gives a detectable separation range of $\sim$7-200\,mas. 

The observations were carried out over six half-nights between July $14^{th}$ and August $6^{th}$ 2010. Target stars which were in close proximity to each other on the sky were sectioned into groups of four or five stars. This was done in order to keep constant air mass and seeing conditions between the targets so that later calibration would be made more accurate. This also reduced the time taken to slew between stars during the observation nights. Furthermore, each group of stars was observed twice, with sufficient time between them to allow the Earth's spin to rotate the baseline with respect to the targets and provide a new position angle for the second observation. This allowed a separation on the sky to be found rather than a projection along an individual baseline position angle. Given that we expect to observe new companions with periods of $\sim$1\, year and with orbital motions on the order of $\sim$1$^\circ$ of position angle per day, we have ensured observations are either on the same night or neighbouring nights where orbital motion is insignificant compared to position angle uncertainties. Each observation consisted of recording 100 seconds worth of 3.5\,ms exposures while an interference fringe pattern was locked on the camera. 

\subsection{Data Reduction and Calibration}
\subsubsection{Data Reduction}
The raw image frames recorded by the PAVO camera were reduced into squared visibility values for each of the 25 wavelength channels using a number of IDL programs written by the SUSI group. The pupil image frame was sectioned into an image for each wavelength channel, and these were used to calculate the squared visibility $V^2$. Without going into the fine detail or complexities of SUSI data reduction, the method of calculation is given by Equation \eqref{coolv2}. The pupil is Fourier transformed to yield the power spectrum, and the power of the fringe is totalled and divided by the total flux squared. 

\begin{equation}
V^2 = \frac{\rm{Fringe Power}}{(\rm{Flux})^2}
\label{coolv2}
\end{equation}

Individual frames taken during observations of single targets which showed anomalously low visibilities were rejected based on manual inspection. This is most important for nights where seeing was particularly bad (greater than $\sim$2.5'' \citep{susi_seeing94}), intermittent clouds were present, or technical problems were encountered. The result of the data reduction is a squared visibility in 49 wavelength bands which are interpolated from the 25 wavelength channels observed by the PAVO beam combiner. 

\subsubsection{Calibration}
The visibility profiles provided by SUSI include the influence of various systematics, such as seeing effects, air turbulence in the beam combination enclosure, dust on optical surfaces and response of detectors. These can be removed through calibration against another star which is assumed to have a well characterised point-source-like visibility profile. This is often a star of small angular diameter. The basic assumption is the following:
\\
\begin{equation}
V_{\rm{measured}}^2 = V_{\rm{true}}^2 V_{\rm{system}}^2 ,\\
\label{vsys}
\end{equation}
\\
where $V_{\rm{true}}^{2}$ is the true squared visibility of the target star, $V_{\rm{measured}}^2$ is the measured squared visibility obtained from SUSI observations and $V_{\rm{system}}^2$ is the system response factor (different for each wavelength channel) which must be removed from the data. The calibration is done by taking a star which is assumed to be described by a uniformly bright disk with a diameter that can be predicted by $B-V$ colour and V magnitude. This calibrator star must also be within a few degrees of the target it will calibrate, in order to calibrate for seeing effects. The error in this prediction due to the B-V uncertainty is inherently small because the calibrator diameters are very small (below the resolution limit of the instrument).  A uniform disk model is then fitted to the predicted diameter, producing squared visibilities for each wavelength channel. This predicted visibility profile is taken to be the $V_{\rm{true}}^2$ for the calibrator star, and using Equation \eqref{vsys} the system response $V_{\rm{system}}^2$ is found. Hence, to calibrate a target, the measured visibility profile is divided by the system response found using the calibrator. 

In usual SUSI observing, one or preferably more than one specific calibrator would usually be chosen prior to observation for each science target. However, for the purpose of detecting Sco-Cen binary companions, we have simply used those stars which did not display the characteristic signal of a binary star as calibrators for those that did. This worked well as the observations were done in groups of stars nearby in the sky, and calibrators were hence available nearby on the sky and at a very small time difference (often less than ten minutes). Observed targets with a clear oscillatory squared-visibility profile in the uncalibrated data were set aside and labelled as companion detections. For the remaining observations, many cross-calibrations were manually performed and inspected, allowing subtle detections, good calibrator observations, and suspect data to be identified among the observations. 

Once good calibrators were identified, they were cross-checked with the available literature as a final precaution to ensure that they were not binary or multiple systems, or that they were  multiple systems with companions well outside the SUSI coherence length limit or much fainter than the SUSI detection limits. In general, if a star has a companion with an angular separation greater than $1-2$\,arcseconds, it can still be a valid calibrator. A binary system with the secondary at an angular separation of $\sim$200\,mas has an optical path difference between central fringes which is just beyond than the 30\,$\mu$m coherence length of the SUSI/PAVO beam combiner and thus is not a suitable calibrator. Such a binary system will produce a systematically lower visibility than a corresponding single star. A uniform decrease in visibility across all wavelength channels in the calibrator can be problematic depending on what it is used to calibrate. There is no issue when calibrating an obvious binary which displays more than one visibility oscillation with wavelength, as the astrometry is not affected by a slight mis-calibration, however, in the case of a very-narrow binary ($<$20\,mas separation) a slight shift in $V^2$ up or down can affect the determination of the brightness-ratio of the system. In both cases, the measured separation is not affected. On average, each star with an identified companion had two calibrators with similar airmass, with some having more than two. In a small number of cases only one calibrator was available, though a reliable determination of the system parameters can still be obtained.

\section{Companion Detections}

\subsection{Fitting to the Data}
Once the data have been calibrated a model binary system visibility is fitted to the data. In the fitting, each component in the binary system is treated as a point source. This approach is justified given the colour and magnitude constraints on our sample: we have only selected stars bluer that $B-V=-0.1$ and brighter than $5^{th}$ magnitude in V, placing all our objects firmly in the B-type range. This means that the bluest object $\beta$-Crucis, which has an angular diameter of $\sim$ 0.7\,mas \citep{hanbury74}, is representative of the largest objects observed. This is well below the resolution limit of 7\,mas of the 15 metre baseline at SUSI and hence the binary systems will be observed as two point sources. The equation that was fitted to the visibility profiles was the following;

\begin{equation}
V^2 = \frac{V_p^2 + r^2 V_s^2 + 2rV_p V_s \cos{(\frac{2\pi\vec{B} \cdot \vec{s_b}}{\lambda})}}{(1+r)^2},
\label{binvispoint}
\end{equation}

where $r$ is the secondary to primary brightness flux ratio, $\vec{B}$ is the baseline vector projected onto the sky, $\vec{s_b}$ is the separation of the binary system on the sky  and $\lambda$ is the wavelength of observation \citep{longbase2000}.  $V_p$ and $V_s$ are the primary and secondary star visibility profiles respectively. In the case of perfect system alignment and focus, these would both be equal to unity at all wavelengths (as is the case with point-sources). In order to remove the effects of any de-focus in the beam combination system, we modelled the primary and secondary visibility profiles as Gaussians;

\begin{equation}
V_{p,s} = \exp{(-a (\frac{\vec{B} \cdot s_{p,s}}{\lambda})^2)}
\label{degraded_binfits}
\end{equation}

where $s_{p,s}$ is the separation on the sky of the primary or secondary from the stellar photo-centre, and a is a coefficient to be determined. This adequately models coherence length degradation due to de-focus in the system, leaving close companion observations relatively unaffected and wide separation companions more difficult to detect. To determine the value of $a$ for our system we calibrated against the well characterised $\kappa$-Cen system, which has a $\Delta m = 1.4$ magnitude companion at $\sim$100 milliarcseconds separation. We find a value of $a = 9.5\times10^{-3}$.

\begin{figure*}
\subfloat[$\lambda$ Lup \label{lamlup}]{\includegraphics[width=0.5\textwidth]{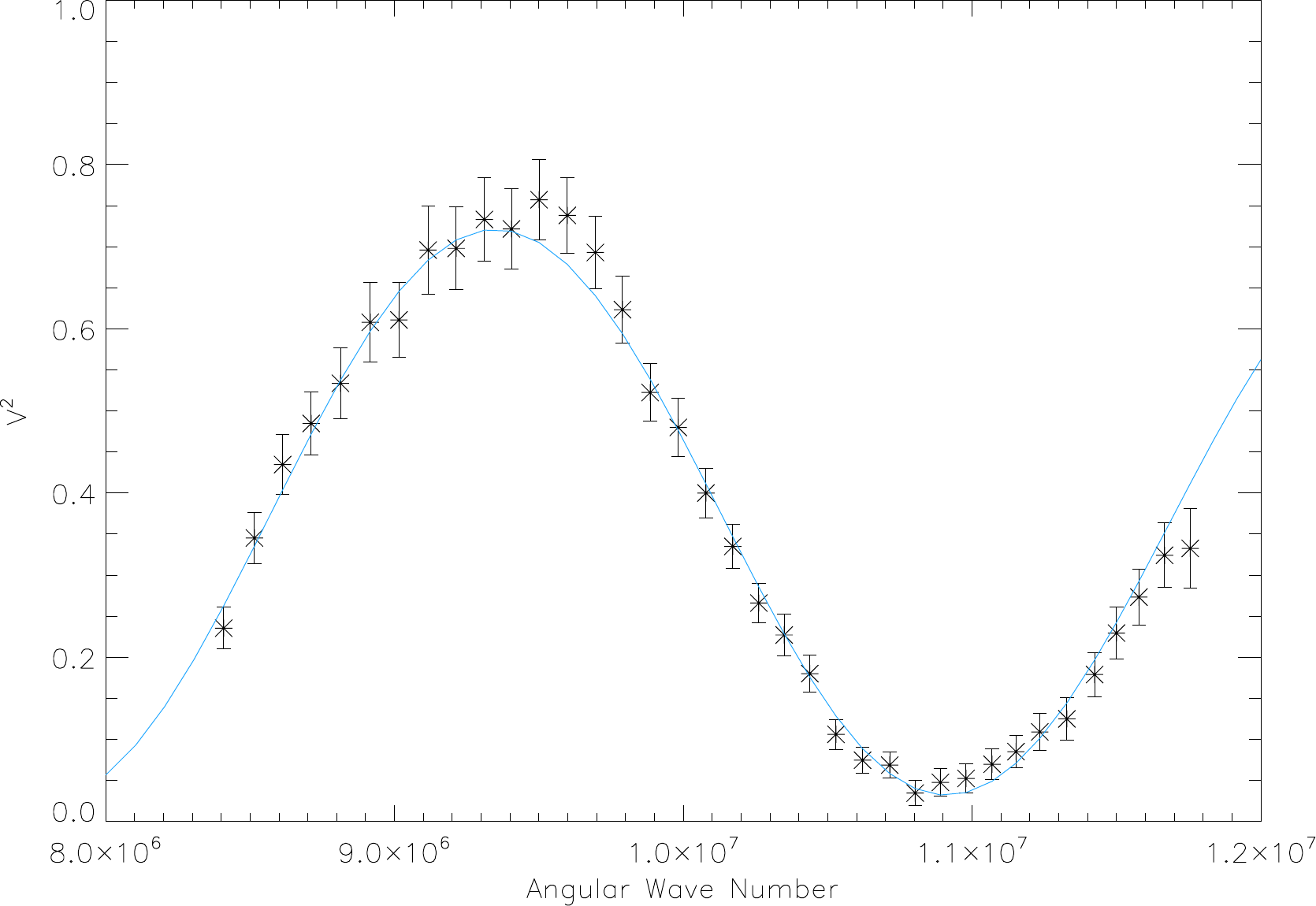}}
\subfloat[$\phi^{2}$ Lup \label{epscen}]{\includegraphics[width=0.5\textwidth]{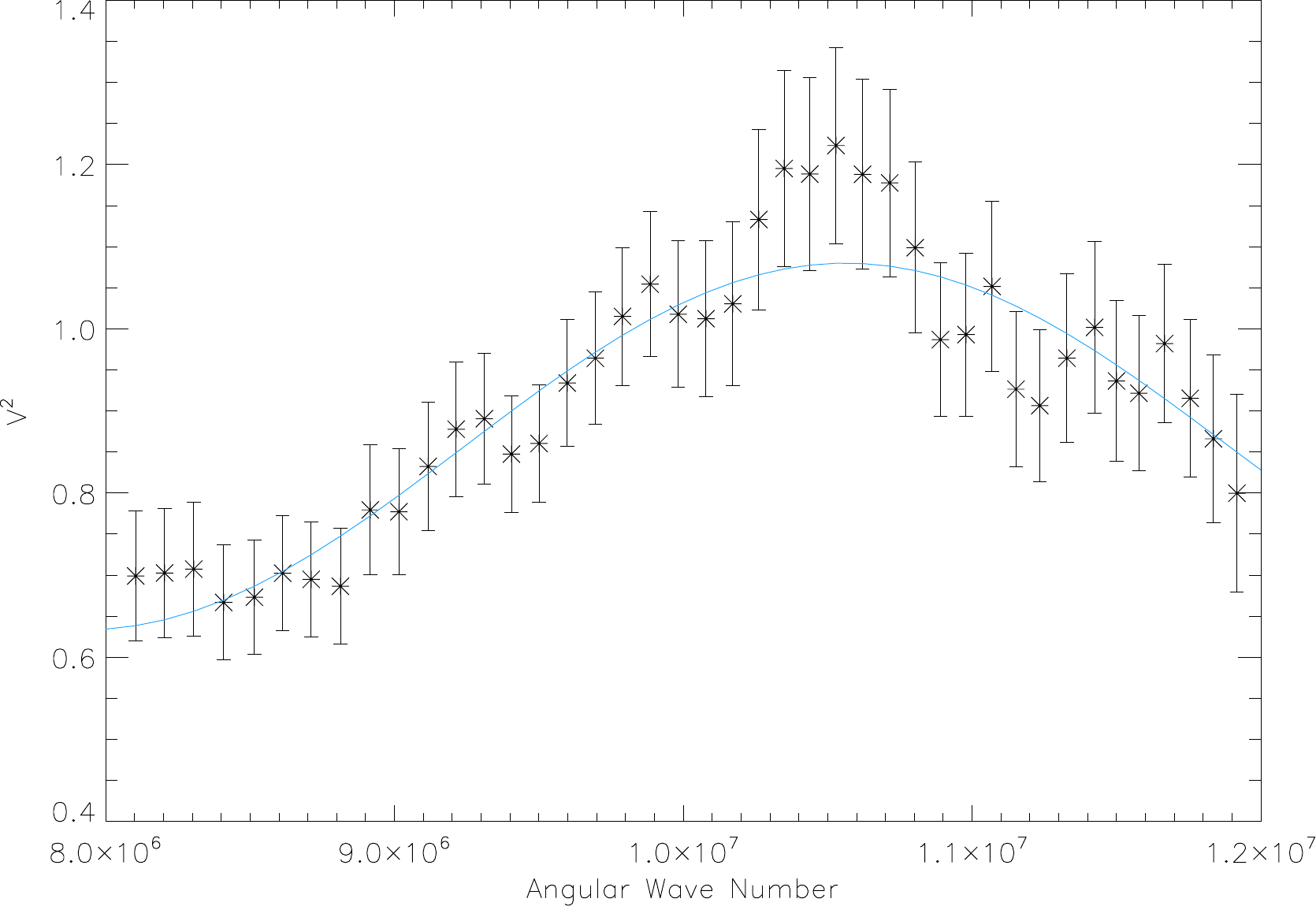}}\\
\subfloat[$\tau$ Sco \label{tausco}]{\includegraphics[width=0.5\textwidth]{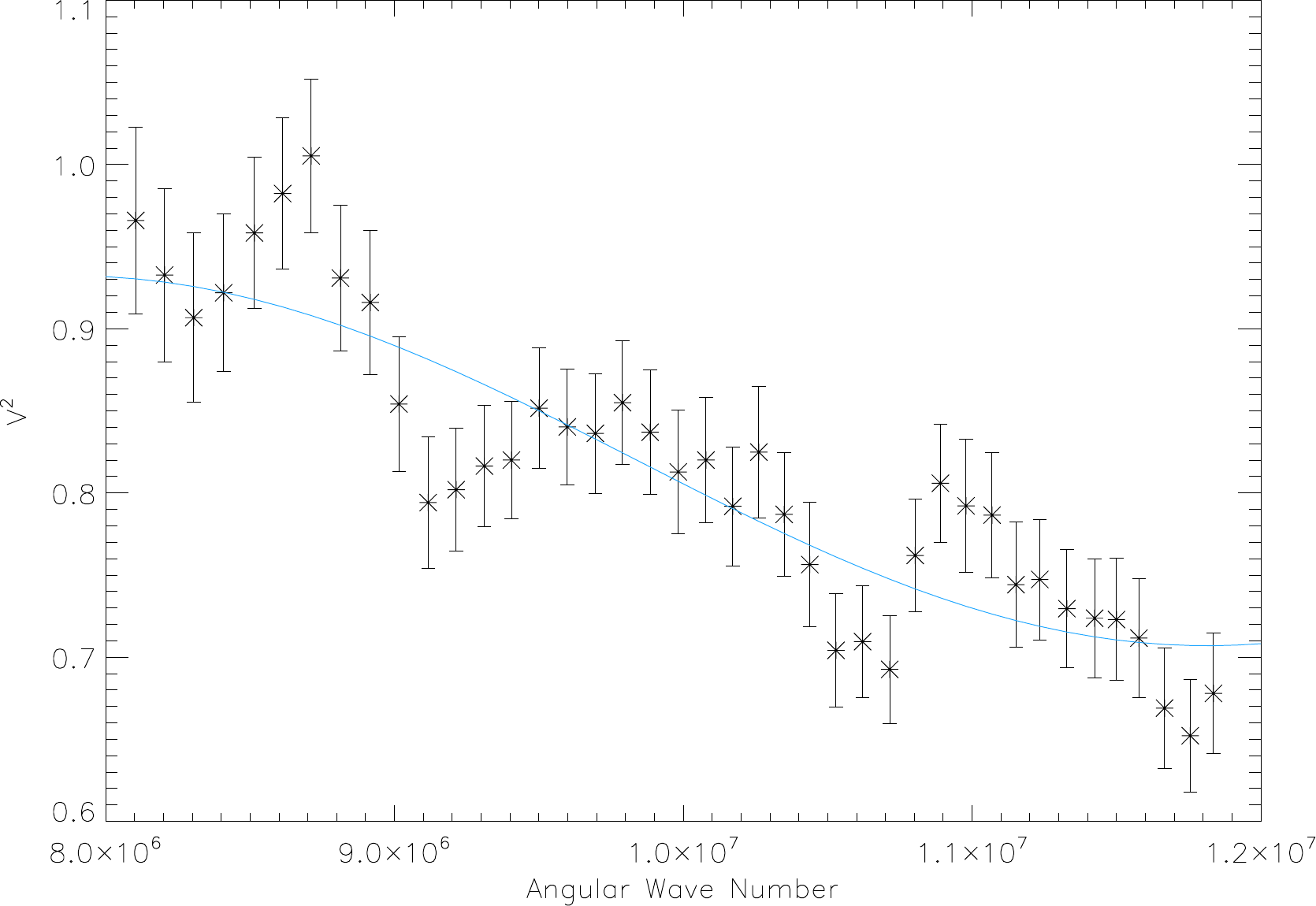}}
\subfloat[$\kappa$ Cen \label{kapcen}]{\includegraphics[width=0.5\textwidth]{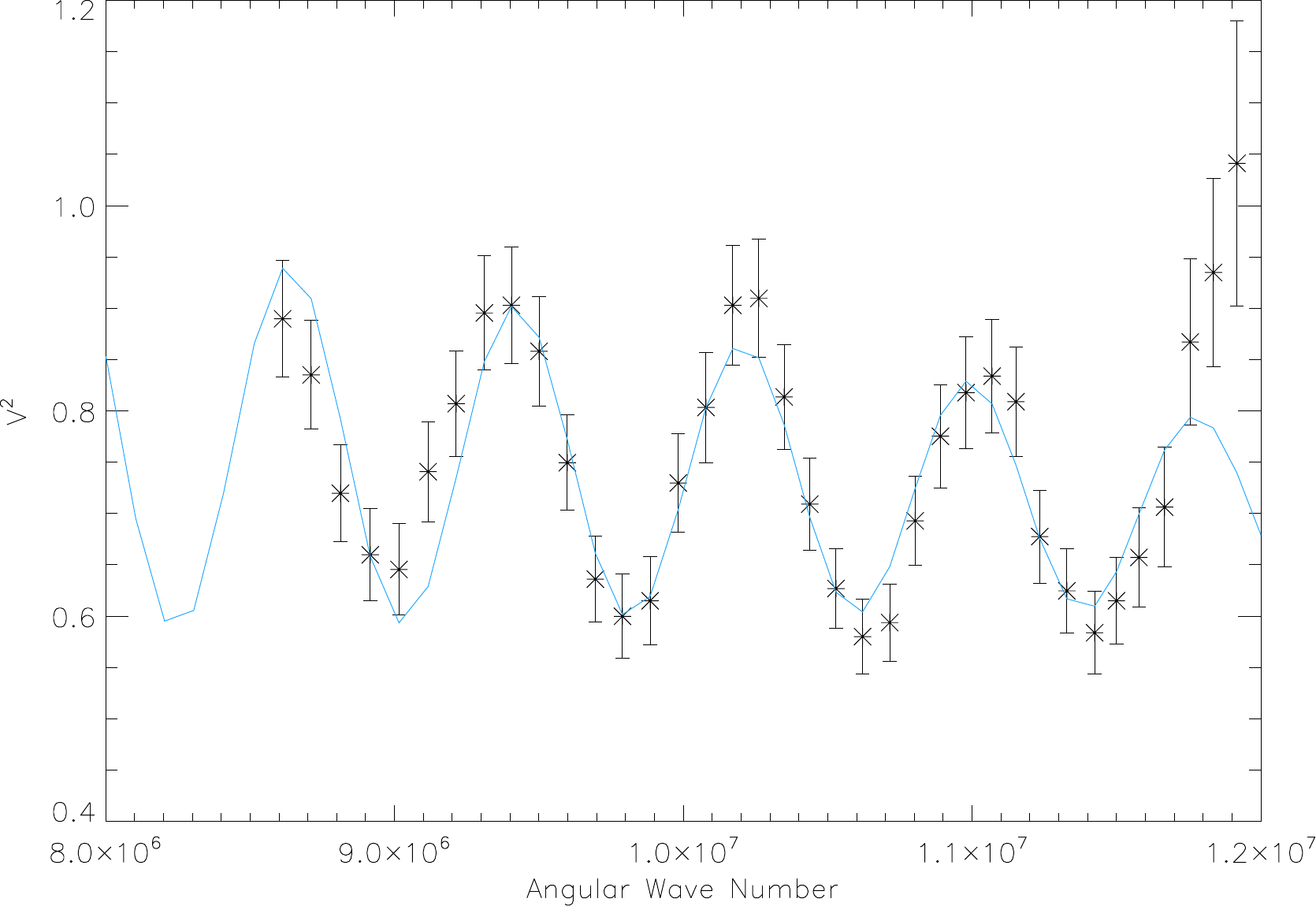}}
\caption{Examples illustrating the typical characteristics of the survey data and the closeness of the binary fits. Figure \ref{kapcen} displays the wide companion against which we calibrated for de-focus. The other three visibility profiles are detections of companions to the stars $\tau$-Sco and $\phi^{2}$-Cen and $\gamma$-Lup. In these figures, the horizontal axis is the angular wavenumber.}
\label{thefits}
\end{figure*}

The fitting process yields both the brightness ratio and the baseline-separation product $(\vec{B} \cdot \vec{s_{b}})$, which is the true separation of the binary system projected onto the direction of the SUSI baseline. Figure \ref{thefits} presents some typical binary visibility profiles and corresponding fits.

With two observations separated by sufficient time, the sky rotates with respect to the baseline and so it is possible to find the true separation of the binary system on the sky at the epoch of observation. The observed separations fitted as described above are in fact the true separations in the north and east directions on the sky under a simple rotation defined by the position angles of observation baseline:

\begin{equation}
R
\begin{pmatrix*}
\rho_N\\
\rho_E
\end{pmatrix*}
=
\begin{pmatrix}
\cos{\theta_1} & \sin{\theta_1}\\
\cos{\theta_2} & \sin{\theta_2}
\end{pmatrix}
\begin{pmatrix*}
\rho_N\\
\rho_E
\end{pmatrix*}
=
\begin{pmatrix*}[r]
\rho_1\\
\rho_2
\end{pmatrix*},
\label{nemat}
\end{equation}

where $\theta_1$ and $\theta_2$ are the position angles, measured North through East, of the two observations, $\rho_1$ and $\rho_2$ are the observed separations and $\rho_N$ and $\rho_E$ are the true separations in the north and east directions on the sky at the epoch of observation. Inverting the matrix R and multiplying on the left gives $\rho_N$ and $\rho_E$. It is important note that there is a $180^\circ$ uncertainty in the position angle of an observed binary system detected using SUSI, this is because a particular binary squared-visibility profile is independent of which star is the brighter component of the system. This mean that a position angle of $0^\circ$ could in fact be $180^\circ$, or that  $\rho_N$ and $\rho_E$ could actually be of the opposite sign. The uncertainty on these two separations can be calculated by transforming the covariance matrix in the standard way:

\begin{equation}
\text{COV} (\rho_{N},\rho_{E}) = 
R^{-1}
\begin{pmatrix*}
\sigma^2_{\rho_1} &0\\
0&\sigma^2_{\rho_2}
\end{pmatrix*}
(R^{-1})^{t},
\end{equation}

where $(R^{-1})^{t}$ is the transpose of $R^{-1}$ and noting that the covariances between the two observed separations are zero because they are completely independent observations. In the cases where more than two observations of a target were done, we used least-squares fitting to calculate the true separation. The observations used in the fits were generally taken on the same night, or over two nearby nights, so that even in the case of the closest and fastest moving companions detected, any orbital motion is insignificant. In the case of the close $\alpha$-Mus companion, which was observed twice in mid July and twice in early August of 2010, we have treated the two nights individually.  
\subsection{Detected Companions and Detection Limits}

Among the 58 Sco-Cen targets we observed, companions were found to be associated with  24 of them, 15 of which are new detections. The fitted parameters, as well as the final combined contrast ratios and separations for each companion can be seen in Table. \ref{bintab}.

\subsection{Detection Limits}
The completeness of this survey is dependent on two parameters, the resolvable range of companion separations and the largest detectable primary to secondary brightness ratio. An upper bound for the former is given by the coherence length of the interferometer as discussed above, and is $\sim$200\,mas, however, this will be reduced by any de-focus in the system. The latter is not obvious directly from the data. Hence we have used a Monte-Carlo scheme to determine detection limits in different binary separation bands. 

\begin{landscape}
\begin{table}
\centering
\caption{The table of observed companion details. Contrast ratios $(\Delta m)$ are in magnitudes, separations ($\rho$) are in milliarcseconds, Obs refers to observation number,  baseline position angles (PA$_B$) are in degrees and $^n$ following a star name indicates a new companion. For each target, if more than one useable observation was taken we also provide a final contrast ratio and the separations in the North and East direction with their full covariance matrix (the final column is the correlation between the two separations), as well as a final, combined separation $(\rho_f)$ and position angle $(PA_f)$.  The high correlations are due to the fact that the position angles differ by much less than 90$^\circ$. Note that the position angles listed have a $180^\circ$ uncertainty, hence, we have chosen to provide all position angles in the $(0 \le PA < 180)$ range.}

\begin{tabular}{ccccccccccccccccccc}
\hline
Name &Date & Obs & PA$_B$($^\circ$) & $\Delta m$ & $\sigma_{\Delta m}$ & $\rho$ & $\sigma_\rho$ &PA$_f$ &$\sigma_{PA_f}$ &$\Delta m_f $& $\sigma_{\Delta m_f}$ &$\rho_f$&$\sigma_{\rho_f}$& $\rho_N$ & $\sigma_{\rho_N}$ & $\rho_E$ & $\sigma_{\rho_E}$ & COR$_{N,E}$\\ 
\hline
4-Lup & 15/07/2010 & 1 & 14.50 & 0.00 & 0.23 & 2.77 & 1.18 &&     &  &  &  &  &     &     &     &     &     \\
$\delta$-Sco & 15/07/2010 & 1 & 179.98 & 2.11 & 0.89 & 87.87 & 0.11 &  &  &  &  &  &  &  &  &  &  &  \\
&15/07/2010 & 2 & 176.67 & 2.11 & 1.04 & 86.61 & 0.17 & 12.34 & 2.22 & 2.11 & 0.02 & 89.95 & 0.80 & 87.87 & 0.11 & 19.23 & 3.57 & 0.54 \\
$\alpha$-Mus$^n$ & 14/07/2010 & 1 & 5.07 & 2.8 & 0.74 & 10.12 & 0.6 &  &  &  &  &  &  &  &  &  &  &  \\
 & 06/08/2010 & 1 & 30.7 & 2.7 & 0.15 & 15.7 & 0.5 &  &  &  &  &  &  &  &  &  &  &  \\
b-Cen$^n$ & 14/07/2010 & 1 & 5.67 & 1.06 & 0.18 & 9.22 & 0.05 & & & & & &  & & & & &     \\
$\beta$-Mus$^n$ & 14/07/2010 & 1 & 6.95 & 3.48 & 0.23 & 18.29 & 0.07 &  &  &  &  &  &  &  &  &  &  &  \\
 & 14/07/2010 & 2 & 13.78 & 3.72 & 0.88 & 13.19 & 0.58 & 120.58 & 3.49 & 3.50 & 0.12 & 45.62 & 4.37 & 23.21 & 0.61 & -39.27 & 4.89 & -0.99 \\
$\delta$-Cen & 15/07/2010 & 1 & 18.23 & 3.45 & 0.87 & 11.63 & 0.89 &  &     &  &  &  &  &     &     &     &     &     \\
$\epsilon$-Cen$^n$ & 15/07/2010 & 1 & 11.62 & 2.59 & 0.33 & 109.11 & 0.20 &  &  &  &  &  &  &  &  &  &  &  \\
 & 15/07/2010 & 2 & 10.50 & 2.54 & 0.30 & 106.53 & 0.18 & 61.83 & 1.86 & 2.56 & 0.03 & 170.50 & 11.22 & 80.48 & 2.60 & 150.31 & 13.37 &   -1.0  \\
$\epsilon$-Lup & 14/07/2010 & 1 & 8.71 & 1.69 & 0.15 & 49.25 & 0.09 &  &  &  &  &  &  &  &  &  &  &  \\
 & 14/07/2010 & 2 & 39.54 & 1.23 & 0.11 & 49.49 & 0.09 & 24.63 & 0.21 & 1.53 & 0.23 & 51.22 & 0.11 & 46.55 & 0.11 & 21.35 & 0.22 & -0.77 \\
f-Cen$^n$ & 26/07/2010 & 1 & 24.34 & 2.34 & 0.56 & 8.44 & 0.37 &  &  &  &  &  &  &  &  &  &  &  \\
& 26/07/2010 & 2 & 28.62 & 1.24 & 0.27 & 8.61 & 0.22 & 41.69 & 21.29 & 2.10 & 0.55 & 8.84 & 2.94 & 6.60 & 2.65 & 5.88 & 5.07 &  -0.99   \\
$\gamma$-Lup & 26/07/2010 & 1 & 9.35 & 32.86 & 0.36 & 62.84 & 0.13 &  &  &  &  &  &  &  &  &  &  &  \\
 & 14/07/2010 & 2 & 8.77 & 1.37 & 0.16 & 59.13 & 0.09 & 89.61 & 0.38 & 2.63 & 0.98 & 371.48 & 15.95 & 2.53 & 2.52 & 371.47 & 15.96 &   -1.0  \\
j-Cen$^n$ & 15/07/2010 & 1 & 30.64 & 2.38 & 0.35 & 40.40 & 0.20 &  &  &  &  &  &  &  &  &  &  &  \\
 & 26/07/2010 & 2 & 26.67 & 3.24 & 0.55 & 45.72 & 0.27 &  &  &  &  &  &  &  &  &  &  &  \\
 & 26/07/2010 & 3 & 32.25 & 3.43 & 0.80 & 36.64 & 0.36 & 145.38 & 7.18 & 3.14 & 0.32 & 95.61 & 10.53 & 78.68 & 6.69 & -54.32 & 11.69 &   -1.0  \\
$\kappa$-Cen & 14/07/2010 & 1 & 10.14 & 1.56 & 0.20 & 113.63 & 0.17 &  &  &  &  &  &  &  &  &  &  &  \\
 & 14/07/2010 & 2 & 5.85 & 1.25 & 0.12 & 110.01 & 0.12 & 31.33 & 1.05 & 1.40 & 0.16 & 121.88 & 1.29 & 104.10 & 0.36 & 63.38 & 2.70 & -0.96 \\
$\kappa$-Sco & 06/08/2010 & 1 & 170.36 & 4.23 & 0.65 & 14.61 & 0.22 & &     & & & &  &     &     &     &     &     \\
$\lambda$-Lup & 26/07/2010 & 1 & 18.36 & 0.93 & 0.06 & 55.14 & 0.04 &  &  &  &  &  &  &  &  &  &  &  \\
 & 27/07/2010 & 2 & 177.07 & 1.73 & 0.06 & 16.94 & 0.04 &  &  &  &  &  &  &  &  &  &  &  \\
 & 27/07/2010 & 3 & 3.53 & 1.25 & 0.04 & 28.75 & 0.02 & 78.31 & 0.14 & 1.49 & 0.23 & 109.87 & 1.21 & 22.25 & 0.17 & 107.60 & 1.24 & -0.56 \\
$\mu$-Cen$^n$ & 14/07/2010 & 1 & 8.50 & 3.15 & 0.37 & 33.13 & 0.14 &  &  &  &  &  &  &  &  &  &  &  \\
 & 14/07/2010 & 2 & 2.99 & 3.22 & 0.53 & 23.36 & 0.19 & 80.20 & 0.21 & 3.17 & 0.04 & 105.55 & 2.40 & 17.96 & 0.30 & 104.01 & 2.46 & -0.92 \\
$o$-Lup & 15/07/2010 & 1 & 16.25 & 0.28 & 0.06 & 42.62 & 0.03 & &     & & &  & &     &     &     &     &     \\
$\phi^2$-Lup$^n$ & 15/07/2010 & 1 & 15.74 & 2.56 & 0.53 & 16.84 & 0.19 &  &  &  &  &  &  &  &  &  &  &  \\
& 15/07/2010 & 2 & 18.81 & 2.05 & 0.17 & 16.72 & 0.08 & 9.94 & 11.77 & 2.17 & 0.26 & 16.92 & 1.03 & 16.67 & 1.21 & 2.92 & 3.63 &  -1.0   \\
$\pi$-Cen & 14/07/2010 & 1 & 6.12 & 1.35 & 0.13 & 26.64 & 0.08 &  &  &  &  &  &  &  &  &  &  &  \\
 & 14/07/2010 & 2 & 11.19 & 0.56 & 0.12 & 34.16 & 0.09 & 78.94 & 0.16 & 1.29 & 0.48 & 90.22 & 1.33 & 17.31 & 0.21 & 88.54 & 1.38 & -0.96 \\
$\rho$-Cen$^n$ & 15/07/2010 & 1 & 19.72 & 1.10 & 0.20 & 54.12 & 0.13 &  &     & & & &  &     &     &     &     &     \\
$\rho$-Lup$^n$ & 14/07/2010 & 1 & 7.28 & 1.28 & 0.07 & 15.40 & 0.07 &  &  &  &  &  &  &  &  &  &  &  \\
& 14/07/2010 & 2 & 6.21 & 1.98 & 0.10 & 15.99 & 0.08 & 123.33 & 5.61 & 1.82 & 0.35 & 35.07 & 4.94 & 19.27 & 0.69 & -29.30 & 5.67 &  -1.0   \\
$\sigma$-Cen$^n$ & 14/07/2010 & 1 & 14.33 & 2.59 & 0.58 & 88.11 & 0.37 & &     & & & &  &     &     &     &     &     \\
$\tau^1$-Lup$^n$ & 26/07/2010 & 1 & 21.88 & 2.60 & 0.32 & 18.84 & 0.11 &  &  &  &  &  &  &  &  &  &  &  \\
& 27/07/2010 & 2 & 1.18 & 2.97 & 0.36 & 18.04 & 0.11 &  &  &  &  &  &  &  &  &  &  &  \\
& 27/07/2010 & 3 & 7.67 & 2.82 & 0.65 & 18.58 & 0.21 & 18.30 & 0.15 & 2.83 & 0.11 & 18.88 & 0.02 & 17.93 & 0.01 & 5.93 & 0.06 & -0.74 \\
$\tau$-Lib$^n$ & 14/07/2010 & 1 & 5.58 & 2.85 & 0.58 & 12.10 & 0.19 & &     &  & &  & &     &     &     &     &     \\
$\tau$-Sco$^n$ & 14/07/2010 & 1 & 12.14 & 2.97 & 0.31 & 11.15 & 0.28 &  &  &  &  &  &  &  &  &  &  &  \\
 & 14/07/2010 & 2 & 52.16 & 2.93 & 0.36 & 20.37 & 0.11 & 70.93 & 0.78 & 2.96 & 0.02 & 21.52 & 0.27 & 7.03 & 0.35 & 20.34 & 0.32 & -0.89 \\

\hline
\end{tabular}
\label{bintab}
\end{table}
\end{landscape}

This was done by creating a sample of synthetic companions to each observed primary, with random contrast ratio and a random separation within the separation band. The faintest synthetic companion detectable at a three-$\sigma$ level (where $\sigma$ is the typical uncertainty in the calibrated squared-visibility profile) is taken as the detection limit. {bf A list of detection limits for each star in a number of separation bands is given in the Appendix.}

\begin{figure}
\includegraphics[width=0.5\textwidth]{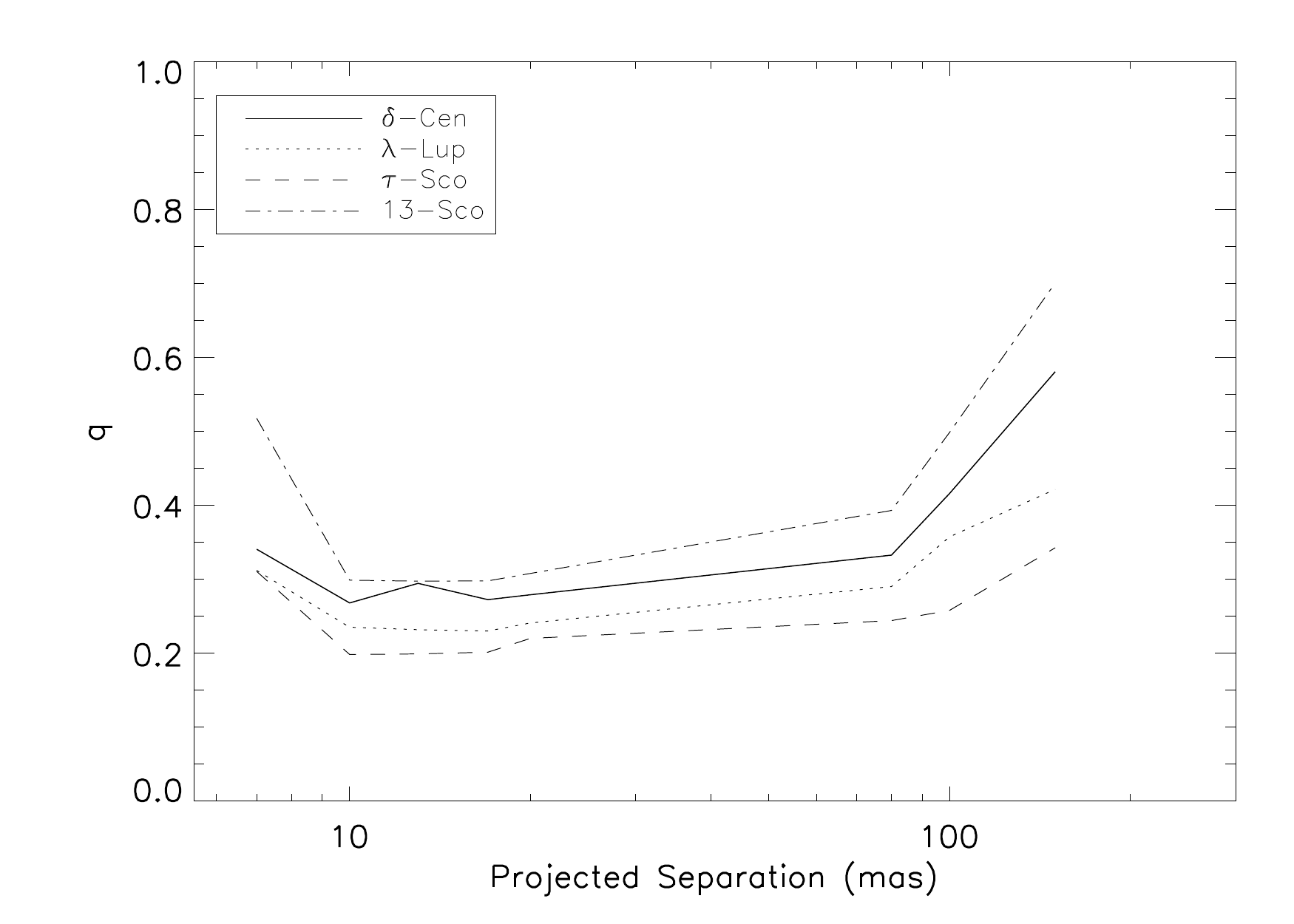}
\caption{Detection limits for four stars observed in our survey.}
\label{det_lim}
\end{figure}

These limits are then converted into mass ratios using isochrones of the mean subgroup ages. Typical detection limits are shown in Figure \ref{det_lim}. 
From the diagram, companions with mass ratios down to typically $q=0.3$ can be detected at separations of $\sim$7-100\,milliarcseconds, at which point the defocus in the system makes wider detections impossible at the smaller mass ratios.

\section{Wide Companions with All-Sky Data}
A primary goal of our study is to create the best possible picture of the multiplicity of the highest-mass stars in the Sco-Cen association. We have moved closer to this goal in the close companion regime with our interferometric survey described above. Conventional and coronagraphic imaging studies complement our work by producing a very complete picture out to $\sim$6\,arcseconds. Beyond these separations, proximity to the primary becomes a rather poor indicator of physical association with the primary. Indeed, any detection beyond $\sim 10^3$\,AU is likely to be a background or foreground contamination. This means the multiplicity catalogs such as the Washington Double Star catalog \citep{wds} are not strictly reliable for separations beyond $\sim$5\,arcseconds. With the availability of all-sky photometry catalogs in numerous bands, such as  2MASS and APASS, it is possible to produce a clearer picture of the wide-separation companion regime.

We undertook a search about our 58 survey targets in the 2MASS \citep{2mass} point source catalog out to a maximum separation of $10^4$\,AU. The 2MASS point source catalog has a resolution of $\sim$5\,arcseconds, meaning that  there will be no overlap between our closer companions and the new companions found here. This searched yielded 670 such possible companions brighter than the $K=14$ 2MASS completeness limit with sufficient near-infrared photometry to allow placement on a colour-colour diagram. We then cross-matched these objects with the APASS \citep{apass} catalog to obtain B and V band magnitudes for the brighter candidates in the sample, and UCAC4 \citep{ucac4} to obtain proper motions. We found 55 of the objects had UCAC4 proper-motions and APASS photometry.
 
We then calculated photometric distances to our companions assuming that they are members of the Sco-Cen association. This was done using Siess isochrones \citep{siess00} of age 6\,Myr for US members and 16\,Myr for members of UCL and LCC . For the brightest candidates, which are expected to fall on the main sequence, we used a Padova main sequence \citep{padova02} to calculate the photometric distances.  Distances were calculated for for ($J-K$,$K$),($H-K$,$K$) and ($B-V$,$V$) where available, and averaged. Photometric distance uncertainties were conservatively estimated to be $\sim$10\%. If a \emph{Hipparcos} parallax measurement was available, this was of course used in place of the photometric distances and uncertainties. A candidate was then deemed a true companion only if the photometric distance and available proper motions showed agreement with the \emph{Hipparcos} proper motion and distance of the primary at the $3\sigma$ level.

We have identified 15 companions in this way, 7 of which had proper-motions, and exclude the other 655 potential companions. The new companions are presented in Table \ref{2masstab}. We note that due to the fact that there are potentially nearby Sco-Cen members to all of the primary stars in our sample, combined with the uncertainty of the photometric distances, there is some chance that association members have been identified as companions. The frequency of spurious companions increases dramatically with the separation, and so we consider companions out to $10^4$\,AU to be reliable, while beyond this limit there is almost certainly significant contamination from other Sco-Cen association members as well as background and foreground objects. 

\begin{table}
\centering
\begin{tabular}{ccccc}
\hline
Primary & Sep ('') & $\delta$K &PA ($^\circ$) & Secondary\\
\hline
$\alpha$-Lup & 25.68 & 6.97 & -127.72 & CD-46 9501B\\
$\beta$-Cru & 42.56 & 7.45 & -34.12 & HD 111123B\\
$\beta$-Mus & 94.78 & 6.87 & 35.36 &\\
$\epsilon$-Lup & 26.29 & 3.85 & 168.70 & CD-44 10066C\\
 & 40.02 & 7.72 & -170.03 &\\
f-Cen & 37.84 & 5.85 & 31.22 &\\
 & 11.53 & 3.43 & 77.82 & HD 113703B\\
$\gamma$-Lup & 53.42 & 10.11 & -67.59 &\\
 & 39.10 & 10.57 & -150.46 &\\
J-Cen & 61.53 & 6.93 & -74.63 &\\
$\kappa$-Sco & 52.84 & 9.75 & -168.68& \\
 & 55.34 & 8.96 & 17.97 &\\
 $\mu^1$-Cru & 35.02 & 0.78 & 17.03 & $\mu^2$-Cru\\
$\mu^2$-Sco & 25.38 & 7.92 & 16.17& \\
$\sigma$-Lup & 26.27 & 6.11 & -156.49& \\

\hline
\end{tabular}
\caption{Companions to our target sample identified from 2MASS and APASS all-sky data.}
\label{2masstab}
\end{table}

\section{Discussion}
\subsection{The Multiplicity Distribution of the Sco-Cen High-Mass Stars}
With the addition of our survey results to the literature, it is possible to study the outcome of multiple star formation among high-mass stars. First we compile all available multiplicity information on the stars in our survey sample from the literature and combine them with our own observations. We then recast the data in terms of separation in astronomical units and mass ratio, rather than angular separations and magnitude differences. We then inspect the distributions of these parameters as a starting-point for a Bayesian analysis of the data, which will provide the most robust determination of the parameters which describe the multiplicity distribution of our Sco-Cen sample. In the Bayesian analysis of the multiplicity distribution which follows, we combine these detection limits with those of \citet{shatsky02}.

\begin{table}[ht]
\centering
\caption{Mass ratios and separations for our sample. Typical uncertainty of the mass ratios is better than 10\%, and the separations in AU for the SUSI companions is typically 10\%. For the wider companions, the uncertainy is of the order of the uncertainty of the 2MASS positions, which are typically 1\% . The source
references are (1) This work, (2) \citet{shatsky02}, (3)
\citet{wds}, (4) This work, all-sky search (5), wide spectroscopic companion, see Table
\ref{specbin_data}, (6) The SUSI study of \citet{tangodelsco}}
\begin{tabular}{cccc}
\hline
Primary & q & $\rho$(AU) & Source\\
\hline
3-Cen & 0.49 & 693.43 & 2 \\
4-Lup & 0.95 & 0.3 & 1, 5\\
$\delta$-Sco & 0.45 & 11.08 & 6 \\
$\alpha$-Lup & 0.08 & 4316.02 & 4 \\
$\alpha$-Mus & 0.01 & 459.23 & 2 \\
 & 0.35 & 2.64 & 1 \\
b-Cen & 0.65 & 1.0 & 1 \\
$\beta$-Cru & 0.06 & 4601.61 & 4 \\
$\beta$-Mus & 0.86 & 95.42 & 3 \\
 & 0.03 & 9044.19 & 4 \\
 & 0.29 & 4.35 & 1 \\
$\delta$-Cen & 0.31 &  12.12 & 1,3 \\
d-Lup & 0.49 & 279.63 & 3 \\
$\epsilon$-Cen & 0.46 & 19.64 & 1 \\
$\epsilon$-Lup & 0.65 & 46.37 & 3 \\
 & 0.19 & 4062.64 & 4 \\
 & 0.02 & 6185.01 & 4 \\
 & 0.64 & 7.92 & 1 \\
f-Cen & 0.21 & 196.83 & 2 \\
 & 0.05 & 4801.44 & 4 \\
 & 0.21 & 1463.15 & 4 \\
 & 0.43 & 1.12 & 1 \\
$\gamma$-Lup & 0.72 & 139.13 & 3 \\
 & 0.01 & 9289.89 & 4 \\
 & 0.01 & 6799.89 & 4 \\
 & 0.39 & 64.61 & 3 \\
HR 4549 & 0.44 & 165.98 & 3 \\
HR 4848 & 0.03 & 850.24 & 2 \\
HR 5543 & 0.13 & 123.20 & 2 \\
j-Cen & 0.03 & 8666.47 & 4 \\
 & 0.32 & 13.47 & 1 \\
$\kappa$-Cen & 0.02 & 661.16 & 2 \\
 & 0.71 & 20.14 & 3 \\
$\kappa$-Sco & 0.01 & 7516.23 & 4 \\
 & 0.01 & 7871.50 & 4 \\
 & 0.26 & 2.08 & 1 \\
$\lambda$-Lup & 0.03 & 82.29 & 3 \\
 & 0.57 & 13.70 & 1 \\
$\mu^1$-Cru & 0.71 & 4052.72 & 4 \\
$\mu^2$-Sco & 0.02 & 4022.14 & 4 \\
$\mu$-Cen & 0.08 & 749.11 & 2 \\
 & 0.33 & 17.05 & 1 \\
$o$-Lup & 0.91 & 5.33 & 1 \\
$\phi^2$-Lup & 0.45 & 3.15 & 1 \\
$\pi$-Cen & 0.59 & 8.89 & 1 \\
$\rho$-Cen & 0.65 & 5.68 & 1 \\
$\rho$-Lup & 0.49 & 3.34 & 1 \\
$\sigma$-Cen & 0.37 & 11.97 & 1 \\
$\sigma$-Lup & 0.06 & 4624.80 & 4 \\
$\tau^1$-Lup & 0.41 & 5.99 & 1 \\
$\tau$-Lib & 0.35 & 1.65 & 1 \\
$\tau$-Sco & 0.30 & 2.84 & 1 \\

\hline

\end{tabular}
\label{bayes_table}
\end{table}

\subsubsection{Compilation of the Sample Data}

In order to produce the most accurate determination of the Sco-Cen multiplicity distribution we have compiled data from a number of sources in the literature. The Bayesian analysis presented below is most easily implemented in terms of mass ratios and physical separations rather than angular separations and magnitude differences. Hence we determined physical separations using the \emph{Hipparcos} parallax measurements \citep{leeuwen07} and the mass ratios using isochrones for the corresponding sub-group ages and the magnitude band used in the original observation. The primary mass was determined using the \emph{Tycho} \citep{perrymantycho} V magnitude and the Padova isochrone of the age of the subgroup which each star is found in \citep{padova02}. The secondary mass was then found by moving fainter along the appropriate magnitude band from the value of the primary. In the case of a small magnitude difference which would not place the secondary in a spectral-type range expected to exhibit premain-sequence (PMS) behaviour the same Padova isochrone was used. For larger contrast ratios which would result in a PMS companion the corresponding Siess PMS isochrone was used \citep{siess00}. The Padova and Siess isochrones show very close overlap (on the order of a tenth of a magnitude) in the higher-mass region of the isochrone in which all Sco-Cen stars observed are past their PMS phase. Any error introduced by this slight difference is, in general, expected to be much smaller than the errors associated with reddening and the measurements of the contrast ratios used to determine the mass ratios, and will not significantly contribute to the outcome of the analysis. The uncertainty of the mass ratios calculated in this way are expected to be typically better than 10\%, which is more than accurate enough for the Bayesian analysis which follows. Similarly, the uncertainty on the physical separations for the SUSI companions are also typically 10\%, while the wider companions in the arcsecond and greater separation regimes are expected to be accurate to $\sim1$\%. Table \ref{bayes_table} provides a list of the calculated physical separations and mass ratios for the non-spectroscopic companions.

For completeness we must also include all spectroscopic companions to the stars in our sample. They provide important information on the smallest separation range of companions and are vital in determining the properties of the multiplicity in the association. We include information on both double and single line companions to our sample stars in two different ways. For the double lined companions there is a directly measured mass ratio for the system, and so the separation can be directly calculated via the orbital period. We have taken the semimajor axes of the binary system and used them along-side the projected separations of our wider companion data. This is justifiable in light of the bin sizes we have used in separation in our analysis and the conversion factors of \citep{dupuy11} which are close to unity for solar-type stars. 
The single line binary companions are not directly useable. The mass ratio and separation of the system cannot be directly determined from the measurements provided by the observation of a single line binary, however they do place useful constraints on the possible values of mass ratio and inclination, and hence also separation, that the systems can have. Table \ref{specbin_data} lists the full information from the literature for the spectroscopic systems in our sample for both the single and double lined systems. 

\begin{table}
\centering
\caption{Spectroscopic companions to stars in our sample, with period ($P$) and mass function ($f(M_1)$. Mass ratios ($q$) are provided for the double lined spectroscopic binaries. The final column lists the literature sources from which the data was taken, they are; (1) \citet{levato87}, (2) \citet{thackeray65}, (3) \citet{thackeray70},(4) \citet{neubauer31}, (5) \citet{uytterhoeven05}, (6) \citet{aerts98}, (7) \citet{cohen08}, (8) This work, (9) \citet{buscombe62}}
\label{specbin_data}
\begin{tabular}{cccccc}
\hline
Star & q & P (days) & $f(M_1)$ & $\sigma_{f(M_1)}$ &   \\
\hline
     3-Cen &         &  17.42800 &    0.00830 &    0.00157& 1\\
     4-Lup &    0.954 &  12.26000 &    0.30680 &    0.03633& 1,2\\
    $\nu$-Cen &         &   2.62528 &    0.00230 &    0.00031 & 1\\
   $\epsilon$-Lup &    0.865 &   4.55959 &  &    & 3,4\\
   $\gamma$-Lup &          &   2.80895 &    0.00073 &    0.00225 & 1\\
   $\tau$-Lib &          0.5&   3.29066 &    0.12626 &    0.04604 &1\\
   $\rho$-Sco &          &   4.00331 &    0.00164 &    0.00050 &1\\
    $\pi$-Sco &            0.78&   1.57010 &    0.27634 &    0.03574 &1\\
 $\xi^2$-Cen &          &   7.64965 &    0.03800 &    0.00322 &5\\
   $\beta$-Cru &    0.625 & 1828.0000 &  &    & 6,7,8\\
    13-Sco &          &   5.78053 &    0.01760 &    0.00410 &1\\
    4-Cen & & 6.930137&0.00598& 0.00143 & 1\\
    e-Lup & & 0.901407& 0.001 & 0.0002 & 9\\
\hline
\end{tabular}
\end{table}

\begin{figure}
\includegraphics[width=0.5\textwidth]{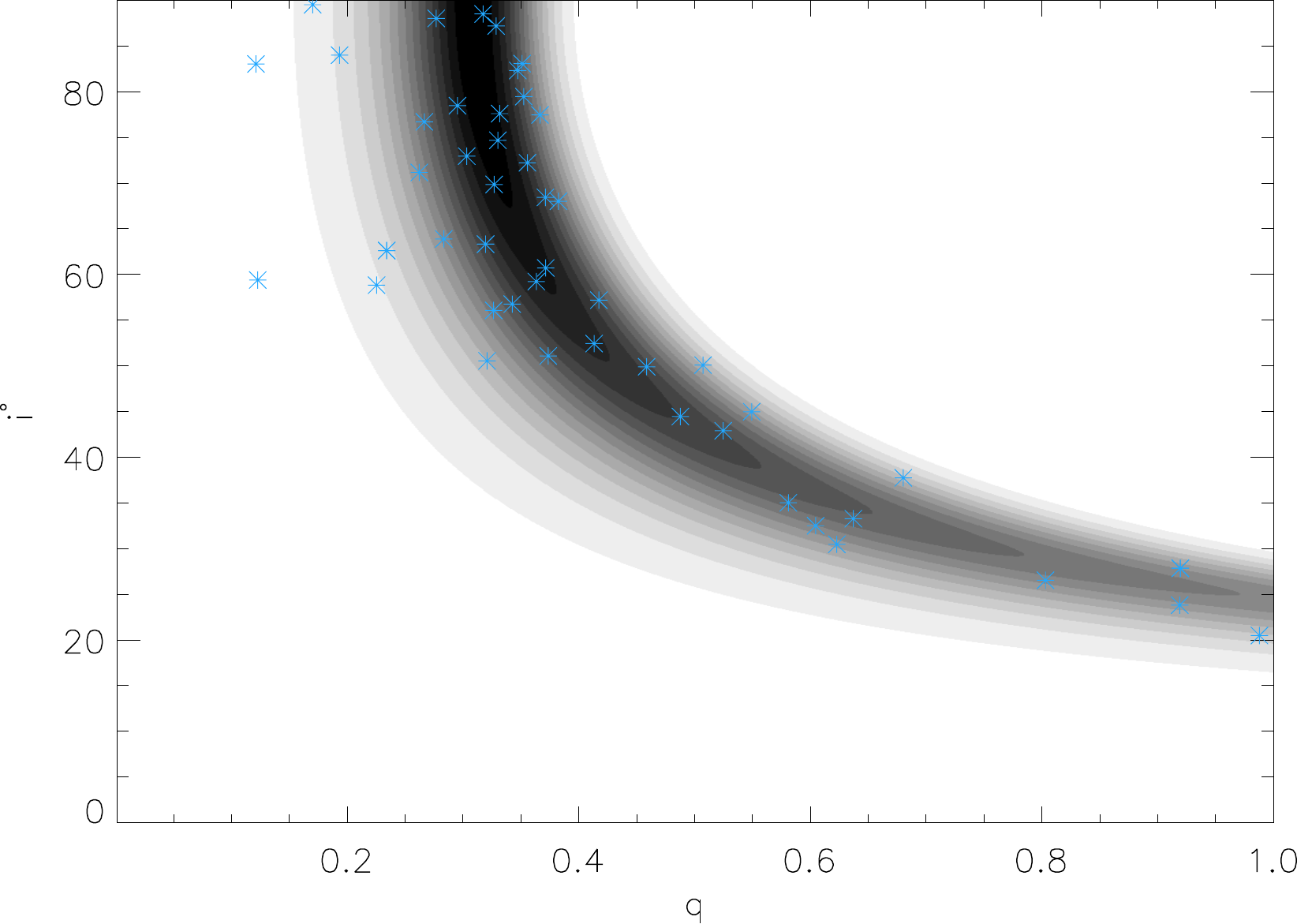}
\caption{An example of the mass ratio-inclination distribution described in equation \ref{specbin_eq}, for the single lined spectroscopic binary system $\gamma$-Lupus \citep{levato87}. The position of the most probable mass ratio and inclination is determined by the observed mass function ($f(M_1)$) of the spectroscopic binary system, while the width of the distribution for any given mass ratio or inclination is determined by the uncertainty in this measurement. The blue points represent a random sampling from the distribution used to represent the ``virtual'' systems used in our analysis.}
\label{specbin_dist}
\end{figure}

We deal with the unknown mass ratio and separation of the single lined systems in the following way: Firstly, we use the observed mass function $f(M_1)$ of the system to determine the distribution of possible values of mass ratio and inclination, based on a primary mass taken from the spectral type and colour of the stars and the corresponding Padova isochrone. Bayes' theorem states the following for the case of a single lined binary system;

\begin{equation}
P(q, i|f(M_1)) = \frac{P(f(M_1)|q, i)P(q, i)}{P(f(M_1))},
\label{bayes_specbin} 
\end{equation}

where $M_1$ and $q$ are the mass of the primary and the secondary to primary mass ratio respectively, and $P()$ denotes probability. We interpret this by first treating the probability of the observed mass function value as unity, ($P(f(M_1))=1$), because we will use the uncertainty in the measurement in the calculations of $P(f(M_1)|q, i)$. $P(q, i)$ is the prior probability distribution of mass ratio and inclination of the orbit. The mass distribution of the companions is unknown, and is one of the properties we wish to determine, hence we define it as uniform up to a mass ratio of 1, and zero beyond it. The distribution of inclinations ($i$), for purely geometric reasons, follows a sinusoidal distribution between 0 and $\pi/2$ radians, if the handedness of the orbit is not considered. For our purposes, treating clockwise and anticlockwise orbits as identical  will not affect the outcome of our analysis, as we only require masses and separations. Hence it is defined as $P(q, i) = \sin{i}$. Finally, we define $P(f(M_1)|q,i)$ to be a gaussian with mean given by the observed mass function of the system and standard deviation defined by the uncertainty in the mass function measurement;

\begin{equation}
P(q, i|f(M_1)) = \frac{\sin{i}}{\sqrt{2\pi}\sigma_{f}}\exp{\big(-\frac{(f(M_1)-f_{mod}(M_1))^2}{2 \sigma_{f}^2}\big)},
\label{specbin_eq}
\end{equation}

where $f_{mod}(M_1)$ is the ``model'' mass function calculated from a given value of mass ratio ($q$) and inclination ($i$). This produces a probability distribution similar to Figure \ref{specbin_dist}. The distribution shows that for each mass ratio $q$ there is a clear range of allowable inclinations which can produce a mass function which agrees with that given by the observations. The position of the allowable mass ratio-inclination pairs is determined by the observed mass ratio and the estimated primary mass. Historically, at this point an expected value of inclination can be chosen, however, this would not represent the observations as closely as possible. The optimal approach is to generate a sample of ``virtual'' systems for each observation based on the described probability density functions (PDFs). We do this by sampling from the described PDF for each system using rejection sampling, which maps a random uniform distribution onto an arbitrary PDF. We take 30 samples for each single lined binary system and include all of these ``virtual'' systems in our sample. 

It is important to note that, while the above method of dealing with single lined spectroscopic binaries is an improvement on simply choosing an expected value of $\sin{i}$ such as 0.8, it is nevertheless still invariably tangled with prior assumptions. Primarily, we have used a range of allowed values as a substitute for the true value, and this has the potential to bias further results. Despite this, our analysis will still produce a robust estimate of the multiplicity properties of Sco-Cen.

Combining both the visual and spectroscopic companions we have a complete picture of the state of knowledge on the multiplicity of the stars in our sample. With this we can determine the multiplicity characteristics of the population of the highest mass stars in Sco-Cen. 

\begin{figure}
\subfloat[\label{qhist}]{\includegraphics[width=0.5\textwidth]{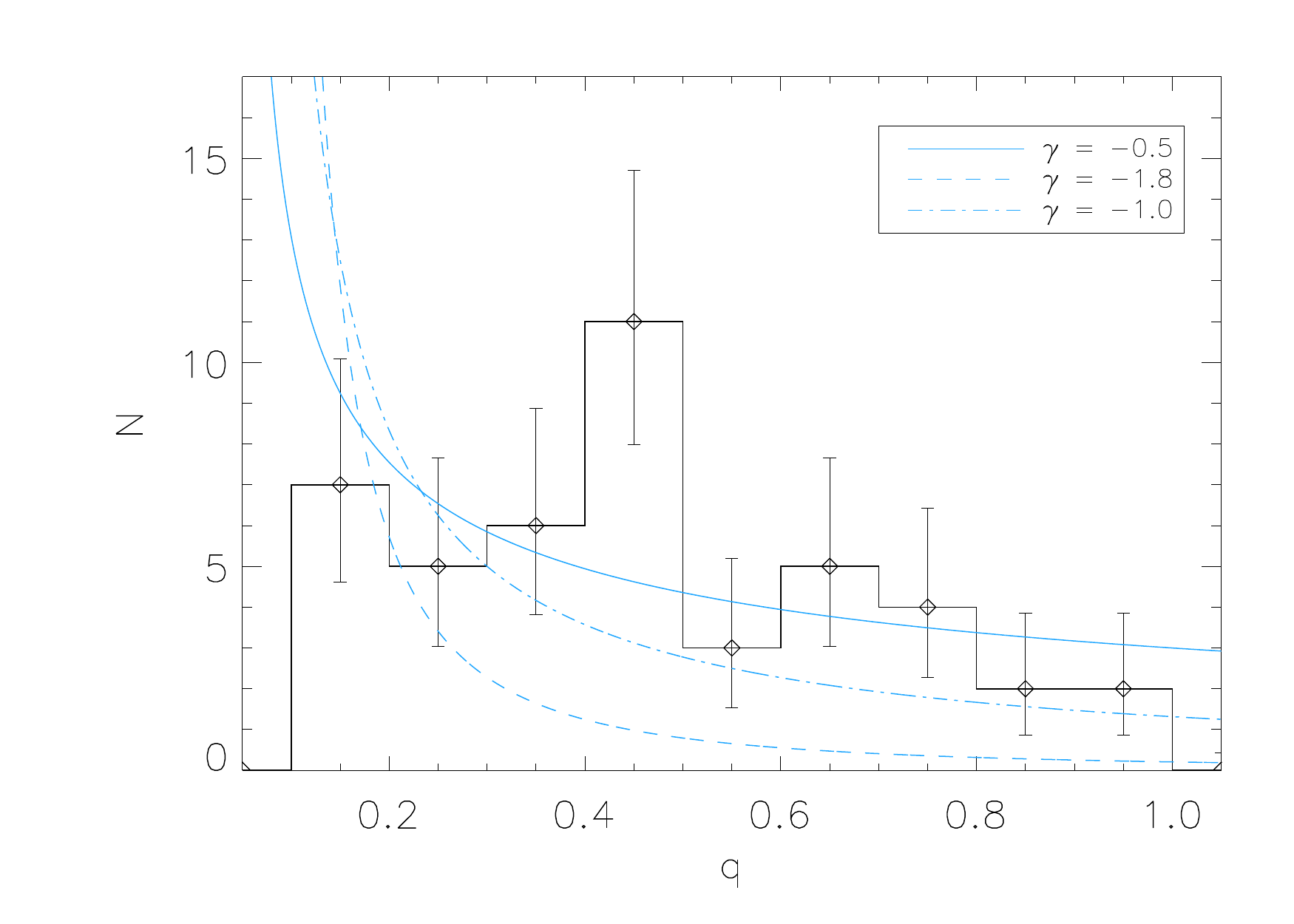}}\\
\subfloat[\label{sephist}]{\includegraphics[width=0.5\textwidth]{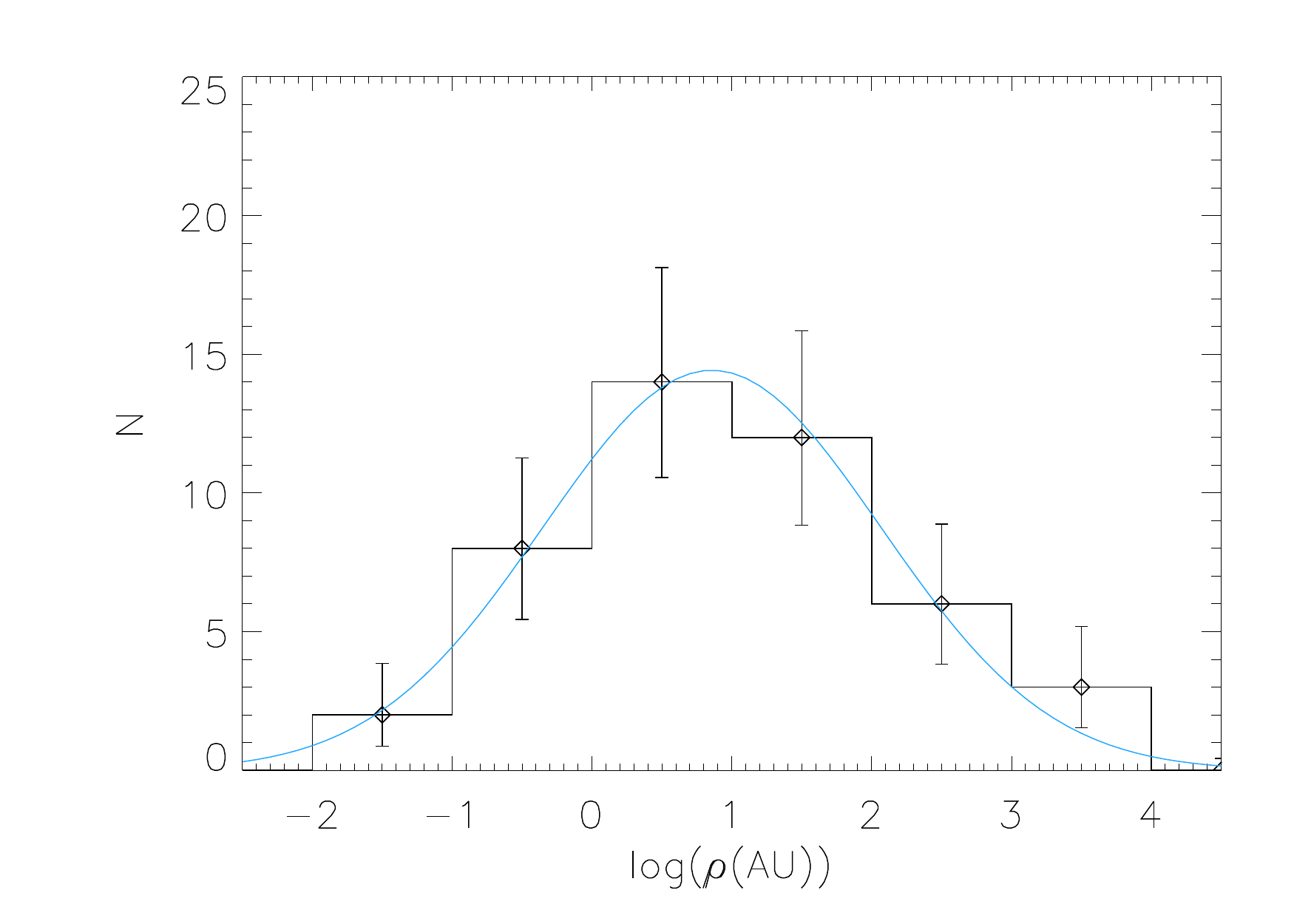}}\\
\caption{Simple histograms displaying the mass ratio and separation (AU) of the companions for the stars in our sample with $q>0.1$. The most likely values of the spectroscopic binary parameters from the PDF's were taken for inclusion in this plot. The mass ratio in the first figure appears to follow a negative power-law distribution with exponent of approximately $-0.5$, and the separation of the companions in the second follows a lognormal distribution with mean of $\sim 0.9$ and spread of $\sim 1.28$. The blue lines in the first figure illustrate mass ratio distributions with different power-law exponents, and in the second figure represent the best fit log-normal Gaussian distribution.}
\label{simple_histograms}
\end{figure}

\subsubsection{Bayesian Analysis}
Classically, the standard method of illustrating binary population statistics is to create histograms of the important quantities, such as separation and mass ratio within the completeness limits of the available data. A model is then fit to the histograms to derive the population parameters. This approach is most useful when the functional forms of the distributions are completely unknown. When a functional form can be determined, a more direct and complete method for working with the data is to use Bayesian statistics, where each observation influences a prior PDF. Bayesian statistics, as opposed to histogram fitting procedures, takes into account all available data in an optimal way, which inherently avoids the need for completeness corrections. Bayesian statistics bypasses the step of fitting a distribution to observations by directly yielding the PDF for the model parameters, which is helpful in showing a study's population measurements, and their uncertainty. As stated above, the important requirement in the use of Bayesian statistics is that the analysis can only be used in the presence of some assumed functional forms of the population distributions, meaning that some inspection of the data (usually with histograms) is required as a starting-point for any Bayesian analysis.

Firstly, we present simple histograms to motivate our choice of prior distributions in the Bayesian analysis, these are shown if Figure \ref{simple_histograms}. Figure \ref{qhist} displays the histogram of mass ratios of the companions in our sample with $q>0.1$ and the best fit to the data is shown in blue. We avoid the $q<0.1$ range of mass ratios due to significant unquantified incompleteness which may bias our distribution. In the Bayesian analysis which follows, we treat the $q<0.1$ regime of mass-ratio as unconstrained. Our plot appears to fit a  power-law, and fitting to the histogram gives a best fit exponent of $-0.38\pm0.24$, which agrees with the value of $-0.4$ which \citet{shatsky02} determined to be the most likely distribution based on their K-band imaging data (much of which is included in our sample and analysis below). The distribution of companion separations in our sample ($q>0.1$) is displayed in Figure \ref{sephist}. The data appears to fit a log-normal distribution in separation quite closely, with a mean log-separation in AU of $0.9 \pm 0.2$ and standard deviation of $1.29\pm0.18$. We know there is incompleteness within the sample, in particular, we expect incompleteness in the SUSI separation range (1-10\,AU) below $q \sim 0.2-0.3$ where companions were not always detectable. Beyond 100\,AU the sample can be considered highly complete down to $q=0.1$ with the addition of our all-sky search, and the spectroscopic binary regime is most likely complete, although it is possible that some SB2's were mistaken for SB1's by the early observers.

Given these observed prior distributions, we can use Bayesian statistics to derive the multiplicity parameters of our sample. We again make use of Bayes' theorem;

\begin{equation}
\label{bayes_general}
P(M|D) \propto P(D|M)P(M),
\end{equation}

where D represents the observations or data, M represents some model, namely some set of parameters and assumed functional forms,  which may or may not describe the data, $P(D|M)$ is the probability of obtaining a given observation or data as a function of the model, $P(M|D)$ is called the posterior PDF of the model given the data and P(M) is the prior PDF for the model. Note that both $P(D|M)$ and $P(M|D)$ depend on the model parameters. This framework is applied by starting with the prior PDF and modifying it with an observation, producing a new prior PDF. This new PDF is then used with a subsequent observation to produce a further modified prior PDF, the process is then continued for all available observations. 

The formalism for the application of the above Bayesian statistics to the analysis of multiplicity populations was first introduced by \citet{allen07}, though we present it in a similar way to \citet{kraus11}. The \citet{allen07} method makes use of four parameters: a companion frequency F, a power-law distribution exponent $\gamma$, a mean of a lognormal separation distribution $\log{\rho_m}$ and a standard deviation for the same distribution $\sigma_{\log{\rho}}$. These parameters describe the PDF of the multiplicity population which describes our sample. Each parameter is assigned a prior and the observations are used to modify the priors to yield the population distribution as described above. In our work, we use a similar modification to the companion frequency F as \citet{kraus11}: in our analysis F can be greater than unity, representing the fact that we are dealing with higher order multiple systems and not solely binaries, which is the case in the \citet{allen07} study.

Rather than the observations individually modifying the prior PDF, we group the data into discrete bins of log-separation and mass ratio and compile a function which describes the number of observed companions in each bin, $N_{comp}(q,\log{\rho})$, which is combined with a detection function $N_{obs}(q,\log{\rho})$which describes the number of observations sensitive to a given bin of $q$ and $\log{\rho}$. The detection function is built based on the detection limits of each observation we took, combined with those of \citet{shatsky02}. We then use each set of grouped data  as a single ``observation'' in the Bayesian sense to modify the prior PDF as described above.

The expected frequency of a companion existing in a particular bin of $q$ and $\log{\rho}$ can be easily calculated from the above functional forms using the four parameters;

\begin{dmath}
\label{rfreq}
R(q,\log{\rho}|M) =\\
\Delta q\Delta\log{\rho}\frac{Fq^\gamma(\gamma+1)}{\sqrt{2\pi}\sigma_{\log{\rho}}} \exp{\Big(\frac{-(\log{\rho}-\log{\rho_m})^2}{2\sigma^2_{\log{\rho}}}\Big)},
\end{dmath}

where we have written $M = (F, \gamma, \log{\rho_m}, \sigma_{\log{\rho}})$, the set of model parameters,  for brevity. Hence, for a given number of observations sensitive to a particular $(q,\log{\rho})$ bin, the number of expected companions detected is given by $R N_{obs}(q,\log{\rho})$. From this the value of $P(D|M)$ is described by a Poisson distribution;

\begin{equation}
  \label{pdatmod}
  P(N_{obs},N_{comp}|M) = \frac{(RN_{obs})^{N_{comp}}\,e^{-RN_{obs}}}{N_{comp}!},
\end{equation}

where $M$ once again represents the four parameters describing the expected distributions. We calculate the value of $P(D|M)$ for values of $q$ between 0 and 1 in bins of width 0.1, and for values of $\log{\rho}$ between -2.0 and 4.0 dex, with all bins having width 0.5 dex. We then use the SUSI detection limits to create a map of $N_{obs}$ in different separation and mass-ratio bins. For the spectroscopic binary separation bins, the number of observations $(N_{obs})$ has been scaled to match the number of random samples we took from the single lined spectroscopic binary systems. In our analysis, we treat the mass ratio range of $0-0.1$ as unconstrained to avoid bias due to unknown incompleteness in this regime where detections are often difficult. The results of this analysis will allow quantification of how many stars are missed in this range. Once the probability of each set of parameters in each bin is calculated, we let each value modify the prior distribution as explained above, yielding the posterior PDF.

Given that all of our prior knowledge went into the determination of the expected distribution shapes, we would like to choose priors for our four parameters which reflect a maximum level of ignorance. The companion frequency, F, is a scale independent parameter, and so the most ignorant choice of prior is given by $1/F$ \citep{sivia}. Simirarly the prior for the spread of the separation distribution is given by $1/\sigma_{\log{\rho}}$, as this parameter is also scale independent. Both $\log{\rho}$ and $\gamma$ are completely unconstrained and so we assign uniform priors to them.

\begin{figure*}
  \subfloat[\label{f_mlsep}]{\includegraphics[width=0.47\textwidth]{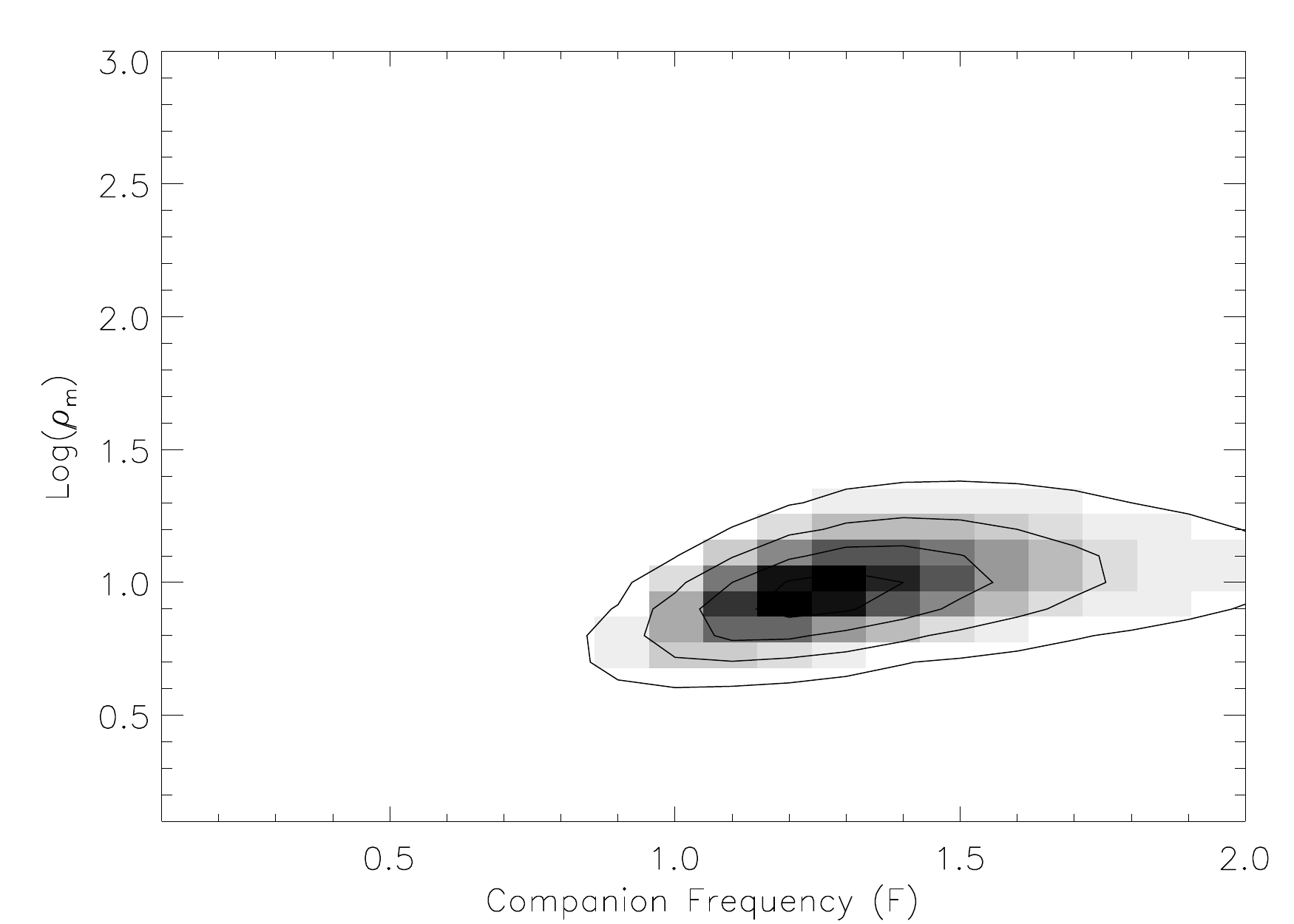}}
  \subfloat[\label{f_slsep}]{\includegraphics[width=0.47\textwidth]{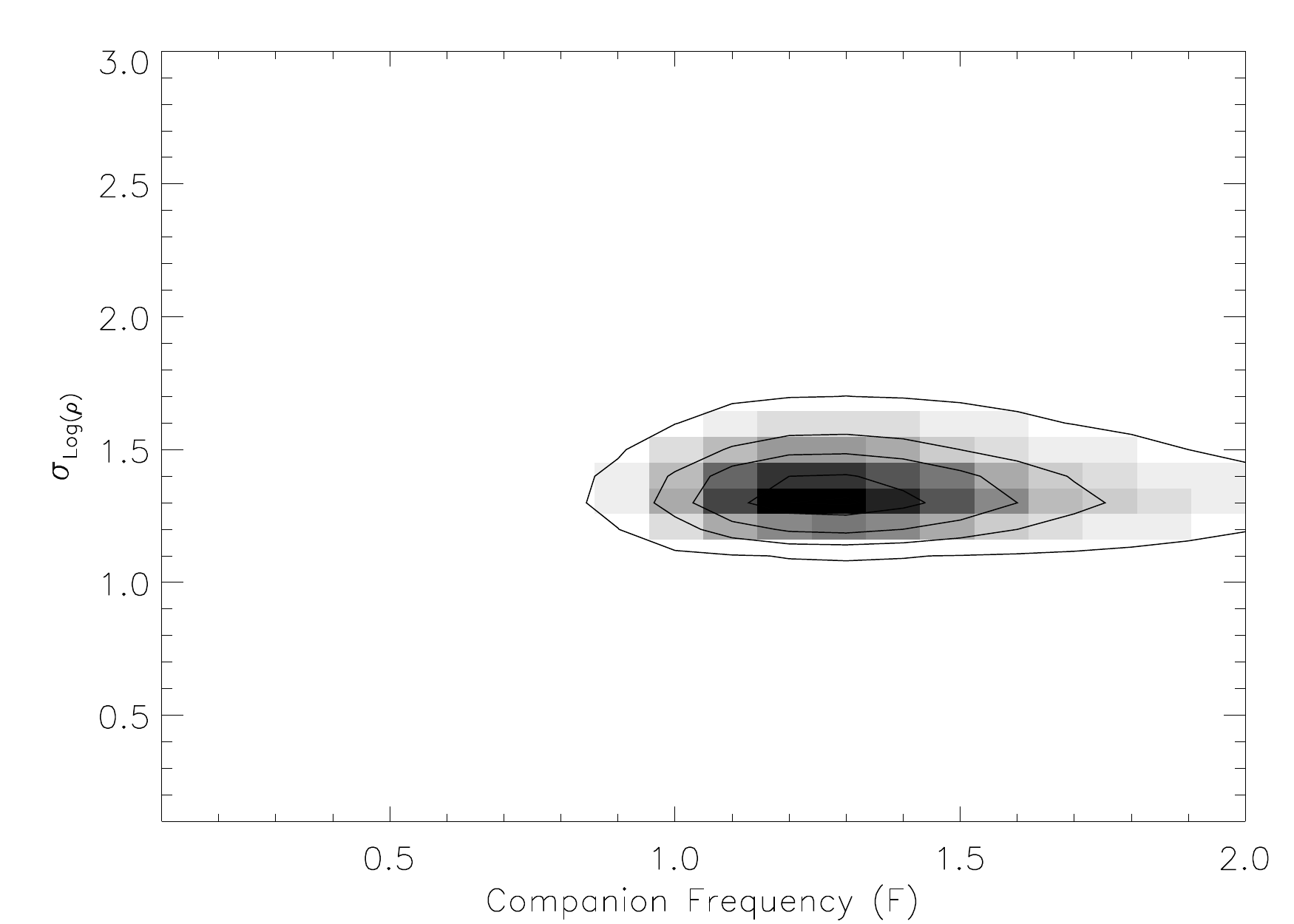}}//
  \subfloat[\label{fline}]{\includegraphics[width=0.47\textwidth]{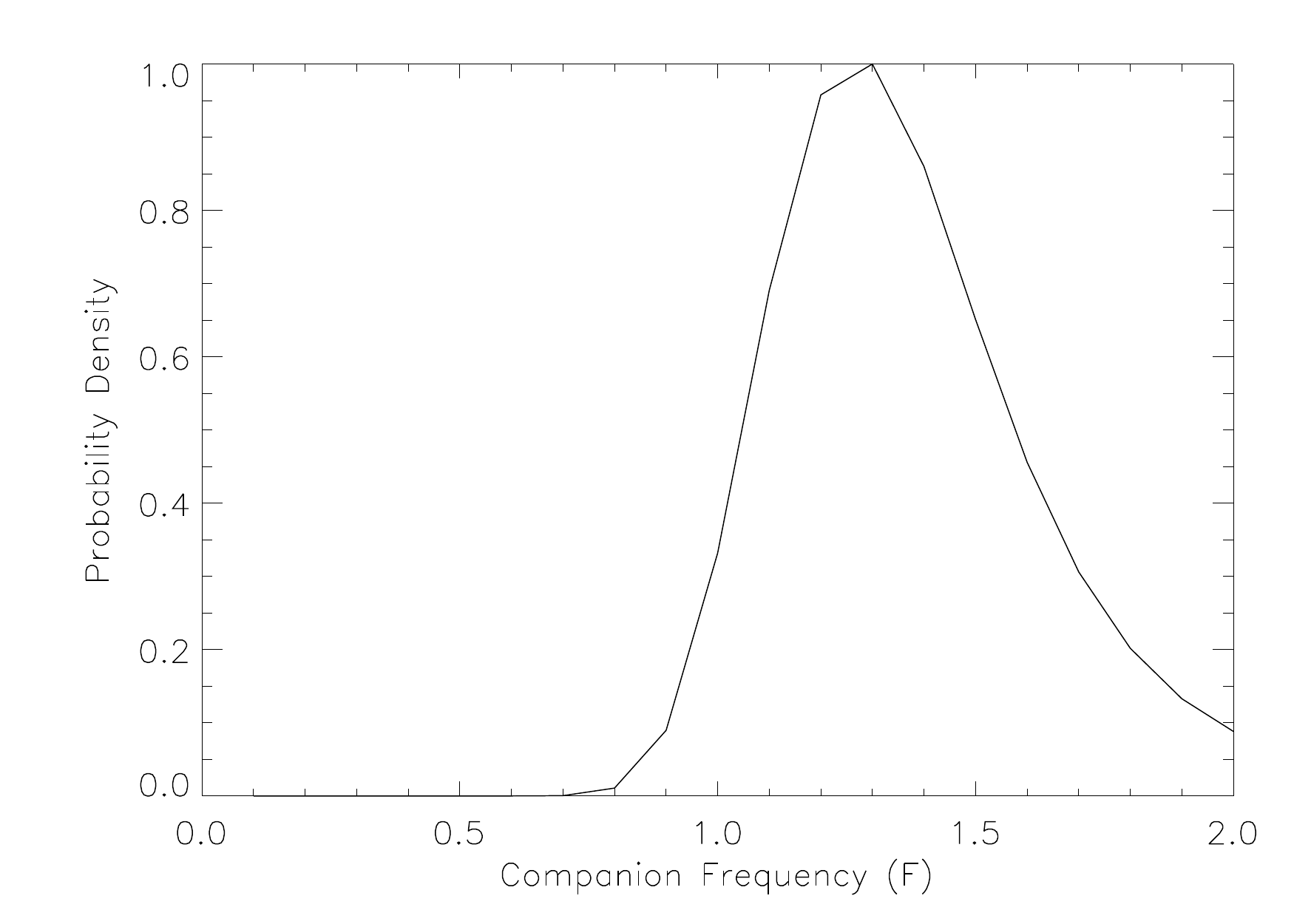}}
  \subfloat[\label{gamline}]{\includegraphics[width=0.47\textwidth]{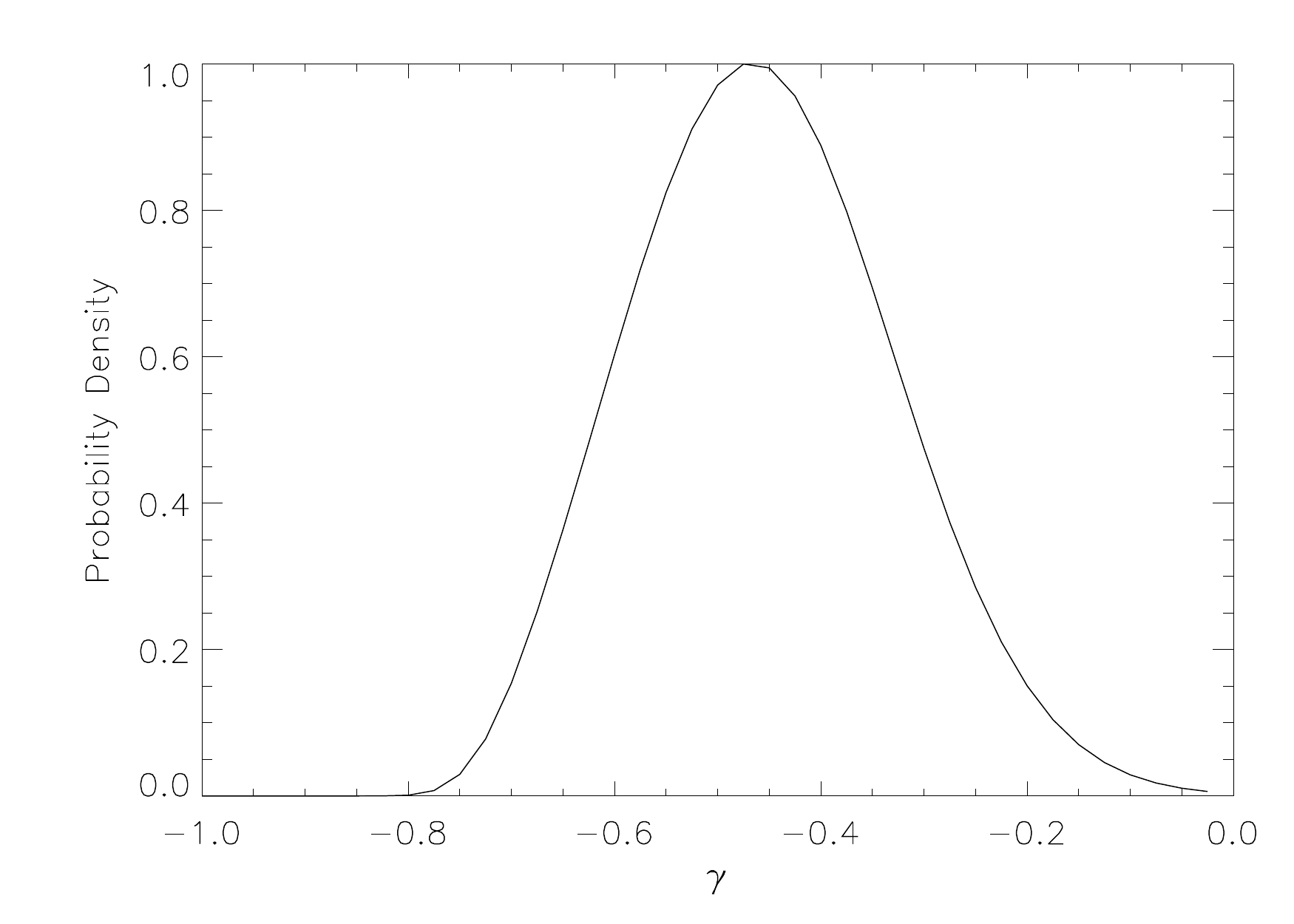}}
  \caption{The marginalised probability density functions produced from our Bayesian analysis in selected correlated dimensions. The figures are as follows: (a) displays the PDF for $F$ and $\log{\rho_m}$ in AU, (b) displays the PDF for $F$ and $\sigma_{\log{\rho}}$, both have contours drawn at 10, 25, 68, 80 and 95\% confidence levels. Figures (c) and (d) display the PDFs for $F$ and $\gamma$ respectively, marginalised over all other parameters and rescaled for ease of display.}
  \label{bayesfigs}
\end{figure*}

The Bayesian analysis we have described here produces a PDF for all possible combinations of the four model parameters and is thus a four-dimensional matrix. To allow presentation of the results, we marginalise the PDF over different sets of parameters and present surfaces and curves for different parameters. The most illuminating results are seen when uncorrelated parameters are shown and others marginalised away. We find that both the companion frequency ($F$) and the mass ratio exponent ($\gamma$) are not correlated with any other parameters, while the $\log{\rho_m}$ and $\sigma_{\log{\rho}}$ are strongly correlated. In Figure \ref{bayesfigs} we have plotted the most useful presentations of the results.

Figures \ref{fline} and \ref{gamline} show very clearly defined peaks for the companion frequency and mass ratio exponent, with values of $F=1.35\pm0.25$ and $\gamma = -0.46\pm0.14$. These results make qualitative sense: The total number of observed companions ($q>0.1$) was 45, hence the vanishing probability of a companion frequency below $\sim$0.8 in Figure \ref{fline}. Our determination of $\gamma$ agrees with the estimated value of $-0.5$ from the \citet{shatsky02} study, although a wider range of possible values is indicated here. The slight difference is not unexpected, as \citet{shatsky02} used only imaging data in their analysis.  Note that the value of the mass ratio exponent $\gamma$ is significantly different for the Sco-Cen high-mass stars compared to that which was determined for lower mass stars in other star-forming regions. The study of \citet{kraus11} found $\gamma$ to be $\sim0$ for 0.25-0.7\,\mdot primaries in the Tau-Aur star forming region, and \citet{allen07} determined a value of $\sim1.8$ for ultra-cool dwarfs. This highlights a potential mass-dependence of the multiplicity outcome of star formation. A further study, using a sample of multiplicity data for the full primary mass range within a single association would further indicate whether this mass trend is present or whether it is related to the specific star-forming regions or associations.

In Figures \ref{f_mlsep} and \ref{f_slsep} we present surface plots of the $F-\log{\rho_m}$ and $F-\sigma_{\log{\rho}}$ PDFs. Both show a clear peak in the PDF at values of $\log{\rho_m}=1.05^{+0.15}_{-0.25}$ and $\sigma_{\log{\rho}} = 1.35^{+0.15}_{-0.25}$. Note the correlation between $\log{\rho_m}$, $\sigma_{\log{\rho}}$ and $F$; a larger value of $\log{\rho_{m}}$ requires larger values of $F$ and $\sigma_{\log{\rho}}$ to account for the number of small separation companion detections. 

\subsection{Single Stars}
The formation of binary or higher order multiple systems is considered as a possible requirement for the conservation of angular momentum in high-mass star formation. Hence, we attempt to ascertain the overall frequency of single stars in our sample. Note that our general result of a companion frequency larger than one, and the large number of companions to stars in our sample are, at least, broadly consistent with the notion that all high-mass stars form with one or more companions. In our sample, there are 17 stars which do not have an observed companion. These stars are listed in Table \ref{singles}.

\begin{table}
\centering
\begin{tabular}{c}
\hline
Single Stars\\
\hline
G-Cen\\
A-Cen\\
$\beta$-Lup\\
$\chi$-Cen\\
$\delta$-Cru\\
$\delta$-Lup\\
$\eta$-Cen\\
$\eta$-Lup\\
$\gamma$-Mus\\
HR 4618\\
HR 5967\\
$\iota$-Lup\\
$\lambda$-Cru\\
$\phi$-Cen\\
$\theta$-Lup\\
$\upsilon^1$-Cen\\
$\zeta$-Cru\\
\hline
\end{tabular}
\caption{The single Sco-Cen stars in the survey sample.}
\label{singles}
\end{table}

The 17 apparently single stars put a hard upper limit of 29\% on the single star fraction among Sco-Cen high mass stars. Using our probability distribution with the most likely parameters determined from the Bayesian analysis, we can estimate the number of single stars which in fact have a companion which was outside of our detection limits by integrating over appropriate separation and mass-ratio regions. We find the most probable number of companions missed in our survey range to be 16, with 4.25$\pm$0.75 of these among the seventeen single stars. We then note that two of the single stars $\beta$-Lup and $\eta$-Lup were not observed by \citet{shatsky02}, leaving the 0.3-5'' arcsecond regime unobserved. From our multiplicity distribution, we expect that 1$\pm$1 of these can have a companion in this separation range. Combining these estimates, this corresponds to an inferred single stars fraction of approximately 17-23\% of the sample. A very simple comparison can be made to our Bayesian model by treating probability of a star having a certain number of companions as a Poisson function with mean given by our most likely value of companion frequency $F=1.35\pm0.25$. This produces a single stars fraction of 20-33\%, and a fraction of quadruple or higher order multiples of 9.5-21\% which is consistent with our single stars fraction and the 8 (12\%) higher order multiples in our sample. The combination of our survey and the literature indicates that there are a number of young high-mass B-type stars which have formed alone, and not as a part of a multiple system.

\subsection{The Effects of Multiplicity on Kinematics}
The effect of multiplicity on kinematics is a significant issue, not just for determining accurate astrometry, but also for understanding how these effects will impact studies using the astrometry. As an example we have calculated a centre of mass (CoM) proper motion for the binary system defined by $\alpha$-Cru A and B. The separations and position angles used to do the calculation were taken from the Washington Double Star catalogue \citep{wds}. $\alpha$-Cru is a wider binary than those observed in our survey, the two measured separations from the catalogue were 5.4 and 4.0\,arcseconds.The position angles were 114 and $112$\,degrees. From the separation and position angle change, the mean motion of the secondary was calculated relative to the primary. This motion was then subtracted from the measured proper motion of the secondary, leaving the CoM proper motion. Our calculated  CoM proper motion was (-36.3,-11.8)\,mas yr$^{-1}$ in right ascension and declination respectively. This is significantly different to the proper motion of the system provided by \emph{Hipparcos} which is $(-35.53,-14.89)\pm(0.45,0.42)$\,mas yr$^{-1}$ \citep{leeuwen07}. Discrepancies such as this which are larger than the typical \emph{Hipparcos} proper motion errors can certainly affect the outcome of, for example, membership selection surveys for moving groups such as Sco-Cen. It is evident that this issue needs to be addressed for a larger sample of wide binaries.

\section{Conclusion}
Our survey of the highest-mass B-type stars in the young Sco-Cen association has determined constraining parameters of 23 companions, and discovered 14 new companions to these stars.

 We used Bayesian statistics and all available multiplicity information to determine the most likely parameters of the multiplicity population of our sample, the results of which agree with previous, less complete analyses. We find that the multiplicity distribution of the stars in our sample to be best described by a log-normal distribution in separation, with a mean of $0.95^{+0.15}_{-0.25}$ and a standard deviation of $1.35^{+0.15}_{-0.25}$, while the mass-ratio follows a power law distribution with exponent $\gamma = -0.47^{+0.13}_{-0.15}$. In addition, the frequency of companions was determined to be $F=1.25\pm0.25$. The multiplicity literature, and our survey results, both point to a very large multiple fraction among high-mass stars in young associations, with only $\sim 17-23$\,\% being single stars according to our statistics. This broadly agrees with the idea that companion formation and companion related mechanisms are the primary angular momentum redistribution method among high-mass stars \citep{larson10}. However, the data suggests asignificant number of single stars among our sample, which according to our Bayesian analysis, are unlikely to fall under the umbrella of missed companions outside of the current detection limits. 
 
Given that the role of magnetic fields in angular momentum loss for high-mass stars is most likely less important, e.g the lack of collimated jets often associated with lower-mass stars \citep{arce07}, some mechanism must be present in the star forming environment which creates single stars. This implies that these stars are either part of a very large-scale wide system, were ejected from a multiple system early in their lifetime, or formed as single stars. Models have suggested that disruptive interactions can shape the formation of high-mass stars in dense clusters  \citep{bonnel03}, but ejection in Sco-Cen is much less likely because OB associations are in general sparse environments. With velocity dispersion on the order of$1^\circ$ per Myr, it is difficult to observationally test ejection hypotheses without GAIA-quality astrometry. The large scale behaviour of Sco-Cen is not completely unknown. It has been shown, using lower-mass members in the \citet{preibisch02} survey of US that two degrees is the approximate wide-scale binarity limit in US \citet{kraus08}. Assuming that UCL and LCC have similar structures,  there is some chance that a small number of the single stars in our sample could be part of a very wide multiple system with one or more other  high mass stars. However, this is unlikely to account for all of the potential  single stars in our sample.  A further possibility is the merger of two lower-mass members of a binary system to form an apparently single, B-type star. While this has been modelled extensively for the case of dense clusters, it is unclear what the frequency of such interactions is in the context of OB associations \citep{zinnecker07, bonnel98}. In all likelihood, the stellar density in Sco-Cen is insufficient to induce binary mergers \citep{bonnelmerger05}.

\bibliography{master_reference.bib}
\bibliographystyle{mn2e}
\appendix
\onecolumn
\section{A complete LIst of Observations}
\label{app1}

\begin{longtable}{cccccccccccc}
\caption{List of observed stars and detection limits ($\Delta m$) in different annular separation ranges in milliarcseconds. The spectral type are taken from the Henry Draper catalogue}\\
\label{obsed_stars}\\
\hline
 Star& HIP & HR & SpT & 7-10 & 10-13  &13-17 & 17-20 & 20-80& 80-100 &100-150 & 150-200 \\
\hline
\endfirsthead
\hline
Star& HIP & HR & SpT &  7-10 & 10-13  &13-17 & 17-20 & 20-80& 80-100 &100-150 & 150-200 \\
\hline
\endhead
\hline
\endfoot
\hline
\endlastfoot

13-Sco  &  79404  &         6028 & B3 &  1.53  &  2.92  &  2.98  &  2.96  &  2.73  &  2.21  &  1.62  &  0.84 \\ 
3-Cen  &  67669  &         5210 & B5 &  2.54  &  3.67  &  3.71  &  3.69  &  3.45  &  2.97  &  2.29  &  1.70 \\  
4-Cen  &  67786  &         5221 & B5 &  1.98  &  3.19  &  3.21  &  3.13  &  2.95  &  2.44  &  1.86  &  1.18 \\  
4-Lup  &  76945  &         5839 & B5 &  1.92  &  3.16  &  3.21  &  3.17  &  2.98  &  2.44  &  1.84  &  1.12 \\  
$\delta$-Sco  &  78401  &         5953 & B0 &  2.25  &  3.57  &  3.74  &  3.66  &  3.49  &  2.85  &  2.24  &  1.60 \\  
G-Cen  &  60710  &         4732 & B3 &  2.68  &  3.22  &  2.90  &  3.29  &  3.06  &  2.66  &  2.17  &  1.45 \\  
a-Cen  &  70300  &         5378 & B5 &  2.20  &  3.14  &  3.11  &  3.11  &  2.91  &  2.48  &  1.86  &  1.30 \\  
$\alpha$-Lup  &  71860  &         5469 & B2 &  2.86  &  2.42  &  2.94  &  3.19  &  3.01  &  2.68  &  2.24  &  1.45 \\ 
$\alpha$-Mus  &  61585  &         4798 & B3 &  4.23  &  3.69  &  4.64  &  4.22  &  4.74  &  4.25  &  3.83  &  3.14 \\  
b-Cen  &  71865  &         5471 & B3 &  2.20  &  3.37  &  3.34  &  3.32  &  3.10  &  2.60  &  1.99  &  1.53\\ 
$\beta$-Cru  &  62434  &         4853 & B1 &  2.29  &  1.97  &  2.33  &  2.61  &  2.47  &  2.12  &  1.60  & 0.85\\  
$\beta$-Lup  &  73273  &         5571 & B2p &  3.08  &  4.05  &  4.04  &  4.09  &  3.79  &  3.24  &  2.65  &  2.11 \\  
$\beta$-Mus  &  62322  &         4844 & B3 &  3.89  &  4.49  &  4.52  &  4.40  &  4.20  &  3.74  &  3.34  &  2.59 \\ 
$\chi$-Cen  &  68862  &         5285 & B3 &  2.11  &  3.10  &  3.20  &  3.08  &  2.91  &  2.41  &  1.79  &  1.21 \\  
$\delta$-Cen  &  59196  &         4618 & B5 &  3.21  &  3.82  &  3.58  &  3.78  &  3.72  &  3.27  &  2.71  &  1.87 \\ 
$\delta$-Cru  &  59747  &         4656 & B3 &  3.38  &  3.03  &  3.44  &  3.68  &  3.52  &  3.12  &  2.69  &  1.90 \\  
$\delta$-Lup  &  75141  &         5695 & B2 &  2.44  &  3.37  &  3.35  &  3.27  &  3.15  &  2.70  &  2.01  &  1.53 \\  
d-Lup  &  76371  &         5781 & B3 &  2.19  &  3.15  &  3.12  &  2.99  &  2.93  &  2.47  &  1.82  &  1.31 \\ 
e-Lup  &  74449  &         5651 & B3 &  1.96  &  3.04  &  3.01  &  2.94  &  2.83  &  2.30  &  1.73  &  1.12 \\  
$\epsilon$-Cen  &  66657  &         5132 & B1 &  3.67  &  4.16  &  4.06  &  4.32  &  4.04  &  3.55  &  3.06  &  2.30 \\  
$\epsilon$-Lup  &  75264  &         5708 & B3 &  3.06  &  3.69  &  3.69  &  3.67  &  3.43  &  3.02  &  2.62  &  1.86 \\ 
$\eta$-Cen  &  71352  &         5440 & B3p &  3.12  &  4.14  &  4.22  &  4.28  &  3.86  &  3.35  &  2.76  &  2.10 \\ 
$\eta$-Lup  &  78384  &         5948 & B3 &  3.24  &  3.96  &  4.01  &  4.04  &  3.76  &  3.20  &  2.64  &  1.98 \\ 
f-Cen  &  63945  &         4940 & B3 &  2.80  &  3.34  &  3.16  &  3.37  &  3.18  &  2.76  &  2.22  &  1.65 \\ 
$\gamma$-Lup  &  76297  &         5776 & B3 &  3.96  &  5.08  &  5.05  &  5.06  &  4.75  &  4.30  &  3.51  &  2.63 \\  
$\gamma$-Mus  &  61199  &         4773 & B5 &  2.48  &  1.81  &  2.98  &  2.18  &  2.92  &  2.61  &  2.29  &  1.66 \\  
... &  57851  &         4549 & B5 &  2.67  &  1.54  &  3.29  &  2.90  &  3.34  &  2.98  &  2.53  &  1.88 \\  
... &  59173  &         4618 & B5 &  3.22  &  3.41  &  3.11  &  3.65  &  3.46  &  3.02  &  2.56  &  1.90 \\  
... &  62327  &         4848 & B3 &  2.92  &  2.59  &  2.98  &  3.23  &  3.09  &  2.70  &  2.28  &  1.56 \\  
... &  72800  &         5543 & B8 &  1.47  &  2.77  &  2.77  &  2.77  &  2.53  &  2.03  &  1.42  &  0.65 \\  
... &  78655  &         5967 & B5 &  1.71  &  2.98  &  3.01  &  3.00  &  2.78  &  2.25  &  1.65  &  0.88 \\  
$\iota$-Lup  &  69996  &         5354 & B3 &  3.13  &  4.06  &  4.04  &  3.98  &  3.88  &  3.29  &  2.68  &  2.06 \\  
j-Cen  &  57669  &         4537 & B5 &  3.22  &  2.82  &  3.33  &  3.54  &  3.38  &  3.05  &  2.53  &  1.90 \\  
$\kappa$-Cen  &  73334  &         5576 & B3 &  2.94  &  3.93  &  3.79  &  3.72  &  3.68  &  3.11  &  2.51  &  1.94 \\ 
$\kappa$-Sco  &  86670  &         6580 & B2 &  4.33  &  5.34  &  5.47  &  5.35  &  5.08  &  4.59  &  3.97  &  2.95 \\  
$\xi^2$-Cen  &  64004  &         4942 & B3 &  3.04  &  3.50  &  3.27  &  3.58  &  3.36  &  2.99  &  2.43  &  1.77 \\  
$\lambda$-Cru  &  63007  &         4897 & B3 &  2.99  &  2.55  &  3.08  &  3.33  &  3.18  &  2.85  &  2.32  &  1.55 \\  
$\lambda$-Lup  &  74117  &         5626 & B3 &  3.19  &  4.18  &  4.28  &  4.33  &  4.01  &  3.41  &  2.84  &  2.24 \\  
$\mu$01-Cru  &  63003  &         4898 & B3 &  2.46  &  1.32  &  2.70  &  2.75  &  2.65  &  2.29  &  1.84  &  1.19 \\  
$\mu$02-Sco  &  82545  &         6252 & B2 &  2.53  &  3.53  &  3.60  &  3.59  &  3.49  &  3.13  &  2.81  &  2.32 \\  
$\mu$-Cen  &  67472  &         5193 & B2p &  2.57  &  3.63  &  3.67  &  3.53  &  3.40  &  2.88  &  2.29  &  1.76 \\  
$\nu$-Cen  &  67464  &         5190 & B2 &  1.37  &  2.85  &  2.91  &  2.85  &  2.65  &  2.14  &  1.54  &  0.00 \\  
$o$-Lup  &  72683  &         5528 & B5 &  2.71  &  3.71  &  3.69  &  3.63  &  3.47  &  2.93  &  2.29  &  1.73 \\ 
$\phi^2$-Lup  &  75304  &         5712 & B3 &  2.17  &  3.34  &  3.39  &  3.32  &  3.12  &  2.61  &  2.03  &  1.32 \\  
$\phi$-Cen  &  68245  &         5248 & B3 &  2.12  &  3.04  &  2.90  &  2.79  &  2.90  &  2.34  &  1.82  &  1.15 \\  
$\pi$-Cen  &  55425  &         4390 & B5 &  2.60  &  3.04  &  2.69  &  3.16  &  3.02  &  2.62  &  2.00  &  1.33 \\  
$\pi$-Sco  &  78265  &         5944 & B2 &  2.93  &  3.35  &  3.41  &  3.58  &  3.62  &  3.39  &  3.09  &  2.52 \\  
$\rho$-Cen  &  59449  &         4638 & B3 &  2.15  &  2.68  &  2.30  &  2.77  &  2.63  &  2.18  &  1.62  &  0.90 \\  
$\rho$-Lup  &  71536  &         5453 & B5 &  2.32  &  3.46  &  3.49  &  3.47  &  3.26  &  2.72  &  2.12  &  1.52 \\  
$\rho$-Sco  &  78104  &         5928 & B3 &  1.95  &  3.24  &  3.31  &  3.35  &  3.08  &  2.59  &  1.93  &  1.33 \\  
$\sigma$-Cen  &  60823  &         4743 & B3 &  2.67  &  2.26  &  2.66  &  3.17  &  3.05  &  2.57  &  2.10  &  1.35 \\  
$\sigma$-Lup  &  71121  &         5425 & B2 &  3.04  &  3.76  &  3.68  &  3.60  &  3.58  &  3.11  &  2.54  &  1.74 \\  
$\tau^1$-Lup  &  70574  &         5395 & B3 &  3.41  &  4.24  &  4.19  &  4.17  &  3.99  &  3.52  &  2.90  &  2.21 \\  
$\tau$-Lib  &  76600  &         5812 & B3 &  2.22  &  3.52  &  3.54  &  3.52  &  3.33  &  2.77  &  2.13  &  1.57 \\  
$\tau$-Sco  &  81266  &         6165 & B0 &  2.91  &  4.01  &  4.00  &  3.97  &  3.75  &  3.49  &  3.35  &  2.68 \\  
$\theta$-Lup  &  78918  &         5987 & B3 &  1.04  &  2.57  &  2.62  &  2.60  &  2.38  &  1.88  &  1.24  &  0.00 \\  
$\upsilon^1$-Cen  &  68282  &         5249 & B3 &  2.13  &  3.11  &  3.00  &  2.83  &  2.88  &  2.42  &  1.75  &  1.17 \\  
$\zeta$-Cru  &  60009  &         4679 & B3 &  2.64  &  2.96  &  2.97  &  3.08  &  2.88  &  2.50  &  1.97  &  1.19 \\ 

\hline
\end{longtable}
\end{document}